\documentclass[twocolumn,showpacs,preprintnumbers,amsmath,amssymb,nofootinbib,floatfix,dvipdfmx]{revtex4}

\usepackage{graphicx}

\def\Journal#1#2#3#4{{#1} {\bf #2}, #3 (#4)}

\def\CPC{Chin. Phys. C}

\def\JCAP{J. Cosmol. Astropart. Phys.}

\def\JCAP{JCAP}
\def\JHEP{JHEP}

\def\MPLA{Mod. Phys. Lett. A}

\def\NPB{Nucl. Phys. B}

\def\PDU{Phys.Dark. Univ.}

\def\PLB{Phys. Lett. B}
\def\PREP{Phys. Rep.}

\def\PLBOLD{Phys. Lett.}

\def\PRL{Phys. Rev. Lett.}
\def\PRD{Phys. Rev. D}

\begin{document}

\title{Asymmetric dark matter and effective number of neutrinos}

\author{Teruyuki Kitabayashi}
\email{teruyuki@keyaki.cc.u-tokai.ac.jp}
\author{Yoshihiro Kurosawa}
\email{4bsnm002@mail.tokai-u.jp}
\affiliation{Department of Physics, Tokai University, 4-1-1 Kitakaname, Hiratsuka, Kanagawa, 259-1292, Japan}
\date{\today}

\begin{abstract}
We study the effect of the MeV-scale asymmetric dark matter annihilation on the effective number of neutrinos $N_{\rm eff}$ at the epoch of the big bang nucleosynthesis. If the asymmetric dark matter $\chi$ couples more strongly to the neutrinos $\nu$ than to the photons $\gamma$ and electrons $e^-$, $\Gamma_{\chi\gamma, \chi e} \ll \Gamma_{\chi\nu}$, or $\Gamma_{\chi\gamma, \chi e} \gg \Gamma_{\chi\nu}$, the lower mass limit on the asymmetric dark matter is about $18$ MeV for $N_{\rm eff}\simeq 3.0$. 
\end{abstract}

\maketitle

\section{Introduction}
\label{sec:introduction}
One of the big motivations for considering the asymmetric dark matter (ADM) scenario \cite{Zurek2014PREP} is that the abundance of baryon $\Omega_b$ and dark matter (DM) $\Omega_{\rm DM}$ is observed to be close to each other $\Omega_{\rm DM} \sim 5\Omega_b$. In an ADM model  \cite{Gelmini1987NPB}, there is a common mechanism that might give rise to the baryon asymmetry as well as the ADM asymmetry and we obtain $\Omega_{\rm ADM}/ \Omega_{\rm b} \simeq (\eta_\chi/\eta_{\rm b})(m_\chi / m_{\rm p})$, where $\Omega_{\rm ADM}$, $\eta_{\rm b}$, $\eta_\chi$, $m_\chi$ and $m_{\rm p}$ are the ADM abundance, the baryon asymmetry, the ADM asymmetry, the ADM mass and the proton mass, respectively. From this relation, the natural scale for ADM is around $5$ GeV, however, mass of the ADM can be as low as a few keV in some models \cite{Falkowski2011JHEP,DAgnolo2015PRD}. Moreover, the light non-asymmetric DM (we call it symmetric DM or simply DM), such as MeV to GeV DM,  remains an elusive blind spot in the current underground searches \cite{Izaguirre2015ArXiv}. 

If there are light ADM particles, the energy density of radiation at big bang nucleosynthesis (BBN) epoch could be changed. It is not necessary that the particle be a dark matter candidate to influence BBN. It only need be a thermal relic, e.g., a particle that was in thermal equilibrium with the standard model particles present when the temperature was compatible with the mass of the relic particle. 

These extra particles contribute to the unknown radiation content of the universe \cite{Steigman2012AHEP,Sorensen2013PASA,Lesgourgues2013}. The energy density of the relativistic particles at BBN (as well as at time of cosmic microwave background (CMB) photons released) is usually expressed in terms of the effective number of neutrinos $N_{\rm eff}$. For the standard cosmology, $N_{\rm eff} = 3$ is expected. Including the effect of slight reheating of the neutrinos from early $e^+e^-$ annihilation, we obtain $N_{\rm eff} = 3.046$ \cite{Mangano2005NPB}. 

The relation between the symmetric DM at BBN and the effective number of neutrinos is extensively studied in the literature \cite{Kolb1986PRD,Boehm2012JCAP,Ho2013PRD1,Ho2013PRD2,Jacques2013PRD,Valentino2013JCAP,Franca2013PRD,Steigman2013PRD,Nollett2014PRD,Serpico2004PRD,Nollett2015PRD,Mirizzi2015PRD,Buen-Abad2015PRD,Heo2015arXiv}. The extra particles at BBN contribute to the effective number of neutrinos in the following two cases:
\begin{enumerate}
\item Direct contribution case: If an extra particle is light enough such as eV-scale sterile neutrino, it is regarded as one of the radiation components. This light particle has contributed directly to the effective number of neutrinos as the so-called dark radiation \cite{Ho2013PRD2,Jacques2013PRD}.
\item Indirect contribution case: Although the extra particles are not light enough to contribute directly to the effective number of neutrinos, its annihilation heats other particles via entropy transfer. Consequently, these extra particles contribute to the effective number of neutrinos indirectly even in the absence of dark radiation. Either increase or decrease of the effective number of neutrinos occurs as follows:

(i) If an extra particle couples more strongly to the neutrinos $\nu$ than to the photons $\gamma$ and electrons $e^-$,  
$\Gamma_{\chi\gamma, \chi e} \ll \Gamma_{\chi\nu}$, its late time annihilation heats the neutrinos more than the photons. Ultimately, this type of extra particle yields an excess of the effective number of neutrinos, $N_{\rm eff} > 3$ \cite{Kolb1986PRD,Serpico2004PRD,Steigman2013PRD,Nollett2014PRD,Nollett2015PRD}. 

(ii) On the contrary, if an extra particle couples more strongly to the electrons and photons than to the neutrinos, $\Gamma_{\chi\gamma, \chi e} \gg \Gamma_{\chi\nu}$, its annihilation heats the electron-photon plasma relative to the neutrino background, leading to a reduction in the effective number of neutrinos below the standard model value, $N_{\rm eff}<3$ \cite{Kolb1986PRD,Serpico2004PRD,Steigman2013PRD,Nollett2014PRD,Nollett2015PRD,Ho2013PRD1}.
\end{enumerate}

The relic density of ADM in the original models is set by the asymmetry around the time of baryogenesis to obtain the single explanation for both baryon and dark matter densities \cite{Nussinov1985PL,Barr1990PLB,Kaplan1992PRL,Kaplan2009PRD}.

Some scenarios to generate primordial dark and baryon asymmetry in the ADM models are proposed such as lepton-number and/or baryon-number violating decay, the Affleck-Dine mechanisms, via phase transition at electroweak baryogengesis, scenarios with dark gauge group and messengers between dark and visible sector (see Ref.\cite{Zurek2014PREP} and references therein). For some of these scenarios, since both DM and anti-DM particles may populate the thermal bath in the early universe, the relic number density of ADM is determined not only by the initial number asymmetry but also the annihilation cross section \cite{Graesser2011JHEP,Iminniyaz2011JCAP,Lin2012PRD}. This asymmetric WIMP framework can accommodate a wide range of dark matter masses and annihilation cross sections. For example, Graesser, et al. applied the asymmetric WIMP scenarios to lepton-number violating and baryon-number violating ADM models \cite{Graesser2011JHEP}. Another example is given by Lin, et al., they reported the model independent constraints on the quantities of light ADM with mass $\sim 1$ MeV - $10$ GeV in the asymmetric WIMP framework \cite{Iminniyaz2011JCAP,Lin2012PRD}. 

The generation mechanism of initial asymmetry, along with ongoing self-annihilation, would be important subject in the study of ADM. We keep the similar mechanism which is proposed in Ref.\cite{Graesser2011JHEP} in mind and we put aside discussion of the particular mechanism to generate the initial asymmetry. We would like to show a model independent analysis in this asymmetric WIMP paradigm.

In the ADM scenarios, the ADM particle $\chi$ and its antiparticle $\bar{\chi}$ have nonzero chemical potentials and the particle $\chi$ is not self-conjugate of the antiparticle $\bar{\chi}$. To study the dependence of the light ADM on the effective number neutrinos, we should take care of the chemical potential of ADM. 

We have a little knowledge about the dependence of the extra particle with nonzero chemical potential on the effective number of neutrinos. For the direct contribution case, a constraint on the chemical potential for a fermionic light DM has been deduced from BBN calculation by Boeckel and S.-Bielich \cite{Boeckel2007PRD}. The correlation between the effective number of neutrinos and the cosmological parameters with light DM particles (including ADM) have been studied by Blennow et al. \cite{Blennow2012JCAP}. These two papers give us many interesting results related to the ADM, however, the dependence of the ADM number asymmetry on the effective number of neutrinos is not clear yet. For the indirect contribution case, there was no study of the relationship between the ADM and the effective number of neutrinos. 

In this paper, the known two methods to estimate the effective number of neutrinos with light extra symmetric DM by Boeckel and S.-Bielich in the direct contribution case \cite{Boeckel2007PRD} and by Steigman in the indirect contribution case \cite{Steigman2013PRD} are slightly extended to the light ADM in a straightforward way. By using the extended methods, we discuss some constraints on a MeV-scale ADM with the effective number of neutrinos in both of direct and indirect contribution cases. Significant lower limit on the ADM mass is obtained in the indirect contribution case. The upper limit on the ADM number asymmetry is also obtained in both cases, however, we observe that this upper limit is not strongly constrained by the relic abundance consideration. 

This paper is organized as follows. In Sec. \ref{sec:asymmetric_dark_matter}, we review the basic picture of the ADM and the effective number of neutrinos. This review does not include any new findings. We would like to present a brief review of the ADM, a detailed analysis of the calculation of the relic abundance of a thermal dark matter candidate that presented in many previously published papers. We also show our notations in this section. 

The new results are reported in Secs. \ref{sec:constraints_direct} and \ref{sec:constraints_indirect}. In Sec. \ref{sec:constraints_direct}, we show the method to obtain a constraint on the chemical potential of ADM in the  direct contribution case which is developed by Boeckel and S.-Bielich \cite{Boeckel2007PRD}. We use their method to obtain the limit on the ADM number asymmetry with the effective number of neutrinos. This is the first result in this paper. In Sec. \ref{sec:constraints_indirect}, we show the useful method which is developed by Steigman \cite{Steigman2013PRD} to estimate the effective number of neutrinos with symmetric DM at BBN in the indirect contribution case. Then, we extend this method to include ADM. This extension is second and main result in this paper. In the same section, the constraints on the mass, number asymmetry and cross sections for ADM are obtained numerically. The actual calculations are important complement to any previously reported results in the literature. These complementary results are the third result in this paper. Finally, Sec. \ref{sec:summary} is devoted to a summary.

\section{Asymmetric dark matter and Effective number of neutrinos}
\label{sec:asymmetric_dark_matter}

\subsection{Relic abundance}
Relic abundance of the ADM has been studied \cite{Scherrer1986PRD,Dolgov1993NPB,Graesser2011JHEP,Iminniyaz2011JCAP,Lin2012PRD,Ellwanger2012JCAP,Gelmini2013JCAP,Baldes2014PRL,Bell2015PRD} based on the methods for the symmetric DM \cite{Steigman1977PLBOLD,Hut1977PLBOLD,Scherrer1986PRD,Srednicki1988NPB,Gondolo1991NPB}. We assume that, at the moment close to the ADM decoupling epoch, the only reactions that change the number of $\chi$ and $\bar{\chi}$ are annihilations and pair creation of $\chi\bar{\chi} \leftrightarrow f\bar{f}$ (there is no self-annihilation and creation such as $\chi\chi \leftrightarrow f\bar{f}$ and $\bar{\chi}\bar{\chi} \leftrightarrow f\bar{f}$ \cite{Ellwanger2012JCAP}). With this assumption in mind, the relic density of $\chi$ and $\bar{\chi}$ is determined by solving the following Boltzmann equations
\begin{eqnarray}
\frac{dn_\chi}{dt}+3Hn_\chi &=& \frac{dn_{\bar{\chi}}}{dt}+3Hn_{\bar{\chi}}
\nonumber \\
&=& -\langle \sigma_{\chi\bar\chi}v\rangle \left(n_\chi n_{\bar{\chi}} - n_\chi^{\rm EQ} n_{\bar{\chi}}^{\rm EQ} \right),
\end{eqnarray}
where $n_\chi$ and $n_{\bar{\chi}}$ denote the number density of $\chi$ and $\bar{\chi}$, respectively. Both $\chi$ and $\bar{\chi}$ may populate the thermal bath (equilibrium) in the early universe. The equilibrium densities $n_\chi^{\rm EQ}$ and $n_{\bar{\chi}}^{\rm EQ}$ in the presence of asymmetry differ by the chemical potential $\mu_\chi$. The detail of the annihilation process is included by the thermally averaged annihilation cross section $\langle \sigma_{\chi\bar\chi}v\rangle$. The Hubble expansion rate is calculated as
\begin{eqnarray}
H=\frac{\pi T^2}{M_{\rm Pl}}\sqrt{\frac{g_\ast}{90}}, 
\end{eqnarray}
during the radiation dominated epoch where $M_{\rm Pl}=2.4\times 10^{18}$ GeV, $g_\ast$ and $T$ denote the reduced Planck mass, the effective relativistic degrees of freedom for the energy density and the temperature of the thermal bath (temperature of the photons), respectively. The effective relativistic degrees of freedom is defined as
\begin{eqnarray}
g_{\ast}(T)=\sum_{i={\rm bosons}} g_i \left(\frac{T_i}{T} \right)^4 + \frac{7}{8}\sum_{i={\rm fermions}} g_i \left(\frac{T_i}{T} \right)^4,
\end{eqnarray}
where  $T_i$ and $g_i$ are the temperature and the number of internal degrees of freedom of species $i$, respectively \cite{Kolb1990,Wantz2010PRD}.

We use the standard definitions
\begin{eqnarray}
x=\frac{m_\chi}{T}, \quad Y_\chi=\frac{n_\chi}{s}, \quad Y_{\bar{\chi}}=\frac{n_{\bar{\chi}}}{s}, 
\end{eqnarray}
where 
\begin{eqnarray}
s=\frac{2\pi^2}{45}g_{\ast s}T^3,
\label{Eq:s}
\end{eqnarray}
is the entropy density and 
\begin{eqnarray}
g_{\ast s}(T)=\sum_{i={\rm bosons}} g_i \left(\frac{T_i}{T} \right)^3 + \frac{7}{8}\sum_{i={\rm fermions}} g_i \left(\frac{T_i}{T} \right)^3,
\label{Eq:gast_s_rela}
\end{eqnarray}
is the effective relativistic degrees of freedom for the entropy density for relativistic particles, which turns out to be
\begin{eqnarray}
g_{\ast s}(T)=\sum_{i={\rm bosons}} g_i F_i^- + \frac{7}{8}\sum_{i={\rm fermions}} g_i F_i^+,
\label{Eq:gast_s_general}
\end{eqnarray}
in a more general case, where
\begin{eqnarray}
F_i^\pm = \frac{45}{4\pi^4}\left(\frac{8}{7}\right)^{\frac{1\pm 1}{2}}x_i^4 \int_0^\infty \frac{y\sqrt{y^2-1}}{e^{yx_i \pm1}} \frac{4y^2-1}{3y}dy,
\label{Eq:Fpm}
\end{eqnarray}
is a function of the particle mass $m_i$ that change smoothly from $F_i^\pm =1$ when the particle is ultrarelativistic  ($x_i=m_i/T \ll 1$) to $F_i^\pm =0$ when it becomes nonrelativistic ($x_i \gg 1$) \cite{Blennow2012JCAP}. 

We assume that the universe expands adiabatically and that $g_\ast$ as well as $g_{\ast s}$ are treated as a constant during the $\chi\bar{\chi}$ annihilation period. In terms of $Y_\chi$, $Y_{\bar{\chi}}$ and $x$, the Boltzmann equations become
\begin{eqnarray}
\frac{dY_\chi}{dx} = \frac{dY_{\bar{\chi}}}{dx}
= -\frac{\langle \sigma_{\chi\bar\chi}v\rangle}{H} \frac{2\pi^2}{45} \frac{g_\ast m_\chi^3}{x^4} \left(Y_\chi Y_{\bar{\chi}} - Y_\chi^{\rm EQ} Y_{\bar{\chi}}^{\rm EQ} \right), \nonumber \\
\end{eqnarray}
and we obtain $d(Y_\chi-Y_{\bar{\chi}})/dx=0$. Thus the ADM number asymmetry, namely, the net comoving densities, 
\begin{eqnarray}
\epsilon=Y_\chi-Y_{\bar{\chi}},
\label{Eq:epsilon}
\end{eqnarray}
is constant. Because the number asymmetry $\epsilon$ is conserved, the ADM asymmetry in equilibrium $\epsilon^{\rm EQ}$ remains at any time. Using the definition of the number asymmetry $\epsilon$ in Eq.(\ref{Eq:epsilon}), we obtain the final form of the Boltzmann equations
\begin{eqnarray}
\frac{dY_\chi}{dx} &=& -\frac{\langle \sigma_{\chi\bar\chi}v\rangle}{H} \frac{2\pi^2}{45} \frac{g_\ast m_\chi^3}{x^4} \left(Y_\chi^2 -\epsilon Y_\chi - Y_\chi^{\rm EQ} Y_{\bar{\chi}}^{\rm EQ} \right),
\nonumber \\
\frac{dY_{\bar{\chi}}}{dx} &=& -\frac{\langle \sigma_{\chi\bar\chi}v\rangle}{H} \frac{2\pi^2}{45} \frac{g_\ast m_\chi^3}{x^4} \left(Y_{\bar{\chi}}^2 +\epsilon Y_{\bar{\chi}} - Y_\chi^{\rm EQ} Y_{\bar{\chi}}^{\rm EQ} \right). \nonumber \\
\label{Eq:BoltzmannEq_final}
\end{eqnarray}

We express the present relic abundance of the particle $\chi$ in the terms of the density parameter $\Omega_\chi$ times the scale factor for the Hubble expansion rate $h=0.673$:
\begin{eqnarray}
\Omega_\chi h^2 = \frac{\rho_\chi}{\rho_{\rm crit}} h^2= \frac{m_\chi s_0 Y_\chi(x\rightarrow \infty)}{\rho_{\rm crit}} h^2,
\label{Eq:Omegah2_chi}
\end{eqnarray}
where $s_0=2.89 \times 10^3$ cm$^{-3}$ and $\rho_{\rm crit} = 3H_0^2/(8\pi G)=1.05h^2 \times 10^{-5}$ GeV cm$^{-3}$ are the present entropy density and the present critical density, respectively. We use the relation of the Hubble expansion rate $H_0 = 100h$ km s$^{-1}$ Mpc$^{-1}$ with the Newtonian gravitational constant $G=6.67\times 10^{-11}$ m$^3$ kg$^{-1}$ s$^{-2}$ \cite{PDG2014}. The present relic abundance of the ADM is calculated to be
\begin{eqnarray}
\Omega_{\rm ADM} h^2 &=& \Omega_\chi h^2 +\Omega_{\bar\chi} h^2 \nonumber \\
&=&2.75 \times 10^8 \frac{m_\chi}{{\rm GeV}} Y_{\rm ADM}(x\rightarrow \infty),
\label{Eq:Omegah2_ADM}
\end{eqnarray}
where
\begin{eqnarray}
Y_{\rm ADM}(x) = Y_\chi(x) + Y_{\bar\chi}(x).
\end{eqnarray}

The observed energy density of the cold dark matter component in $\Lambda$CDM model by the Planck Collaboration is $\Omega_{\rm DM}h^2 = 0.1188\pm0.0010$ (68\% C.L.) \cite{Planck2015arXiv}. 

\subsection{Asymmetry and chemical potential}
The distribution function of particle species $i$ is given by
\begin{eqnarray}
f_i=\frac{g_i}{e^{(E_i-\mu_i)/T_i}+\Theta_i},
\label{Eq:f}
\end{eqnarray}
where $g_i$, $E_i$, $\mu_i$ and $T_i$ denote number of internal degrees of freedom, energy, chemical potential and temperature of particle $i$, respectively. The discrete parameter $\Theta_i$ takes only the following three values: $\Theta_i=+1$ and $\Theta_i=-1$ correspond the Fermi-Dirac distribution and the Bose-Einstein distribution while $\Theta_i=0$ corresponds the Maxwell-Boltzmann distribution. In this paper, we consider a fermionic Dirac type ADM $\chi$ and we take $\Theta_\chi = +1$.

The number density $n_i$, energy density $\rho_i$, pressure $P_i$ and entropy density $s_i$ of particle species $i$ with mass $m_i$ in the isotropic universe are obtained as follows \cite{Kolb1990, Kolb1980NPB}:
\begin{eqnarray}
n_i&=&\frac{1}{2\pi^2}\int_{m_i}^\infty E(E^2-m_i^2)^{1/2}f_i dE,  \label{Eq:n} \\
\rho_i&=&\frac{1}{2\pi^2}\int_{m_i}^\infty E^2(E^2-m_i^2)^{1/2}f_i dE, \label{Eq:rho} \\
P_i&=&\frac{1}{6\pi^2}\int_{m_i}^\infty (E^2-m_i^2)^{3/2}f_i dE, \label{Eq:P} \\
s_i&=&\frac{\rho_i+P_i-\mu_i n_i}{T_i}. \label{Eq:s}
\end{eqnarray}

If the degeneracy of the particle $i$ is small ($\mu_i \ll T_i)$, the Fermi-Dirac distribution can be well approximated by the Maxwell-Boltzmann distribution. In the approximation of Maxwell-Boltzmann statistics, the equilibrium number density of the ADM particle $\chi$ becomes
\begin{eqnarray}
n_\chi^{\rm EQ} = g_\chi \left(\frac{m_\chi T_\chi}{2\pi}\right)^{3/2}e^{(-m_\chi+\mu_\chi)/T_\chi} \left( 1 + \frac{15}{8x}+\mathcal{O}(x^{-2})\right),
\nonumber \\
\end{eqnarray}
for the nonrelativistic limit, $m_\chi \gg T_\chi$ ($x \gg 1$), and  
\begin{eqnarray}
n_\chi^{\rm EQ} = g_\chi \frac{T_\chi^3}{\pi^2} e^{\mu_\chi/T_\chi} \left( 1 - \frac{x^2}{4} + \mathcal{O}(x^4)\right),
\end{eqnarray}
for the ultrarelativistic limit, $m_\chi \ll T_\chi$  ($x \ll 1$) \cite{Kolb1980NPB}. The number density of antiparticle $\bar{\chi}$ is also obtained with the fact that $\mu_\chi = - \mu_{\bar{\chi}}$ in equilibrium. As a result, chemical potential drops out in the product $Y_\chi^{\rm EQ} Y_{\bar{\chi}}^{\rm EQ}$:
\begin{eqnarray}
Y_\chi^{\rm EQ} Y_{\bar{\chi}}^{\rm EQ} = \frac{1}{(2\pi)^3} \left(\frac{45}{2\pi^2}\right)^2 \left(\frac{g_\chi}{g_{\ast s}}\right)^2 x^3 e^{-2x},
\end{eqnarray}
for the nonrelativistic and
\begin{eqnarray}
Y_\chi^{\rm EQ} Y_{\bar{\chi}}^{\rm EQ} = \frac{1}{\pi^4} \left(\frac{45}{2\pi^2}\right)^2 \left(\frac{g_\chi}{g_{\ast s}}\right)^2,
\end{eqnarray}
for the ultrarelativistic cases. The absence of the chemical potential in $Y_\chi^{\rm EQ} Y_{\bar{\chi}}^{\rm EQ}$ simplifies the Boltzmann equations in Eq.(\ref{Eq:BoltzmannEq_final}) in the nonrelativistic and ultrarelativistic cases \cite{Iminniyaz2011JCAP}. 

The chemical potential of the ADM in equilibrium is still surviving on the asymmetry
\begin{eqnarray}
\epsilon^{\rm EQ}=Y_\chi^{\rm EQ}-Y_{\bar{\chi}}^{\rm EQ}=\frac{1}{s}\left(n_\chi^{\rm EQ}-n_{\bar{\chi}}^{\rm EQ}\right).
\end{eqnarray}
For the examples, we obtain
\begin{eqnarray}
\epsilon^{\rm EQ} = \frac{45}{2\pi^2} \frac{g_\chi}{g_{\ast s}} \left(\frac{m_\chi}{2\pi T_\chi} \right)^{3/2}e^{-m_\chi/T_\chi} \left( e^{\mu_\chi/T_\chi} - e^{-\mu_\chi/T_\chi} \right), \nonumber \\
\end{eqnarray}
for the nonrelativistic \cite{Gelmini2013JCAP} and 
\begin{eqnarray}
\epsilon^{\rm EQ} = \frac{45}{2\pi^4} \frac{g_\chi}{g_{\ast s}}  \left( e^{\mu_\chi/T_\chi} - e^{-\mu_\chi/T_\chi} \right),
\end{eqnarray}
for the ultrarelativistic cases. From these equations, we have \cite{Gelmini2013JCAP}
\begin{eqnarray}
\frac{\mu_\chi}{T_\chi} = \ln\left[ \frac{1}{2} \left(\frac{\epsilon^{\rm EQ}}{\lambda} + \sqrt{\left( \frac{\epsilon^{\rm EQ}}{\lambda} \right)^2 + 4}\right)\right],
\end{eqnarray}
where
\begin{eqnarray}
\lambda=\frac{45}{2\pi^2} \frac{g_\chi}{g_{\ast s}} \left(\frac{m_\chi}{2\pi T_\chi} \right)^{3/2}e^{-m_\chi/T_\chi},
\end{eqnarray}
for the nonrelativistic and 
\begin{eqnarray}
\lambda=\frac{45}{2\pi^4} \frac{g_\chi}{g_{\ast s}},
\end{eqnarray}
for the ultrarelativistic cases.

We observed the ratio $\mu_\chi/T_\chi$ in many equations. This ratio is the so-called degeneracy parameter or the pseudochemical potential. In the remaining part of this paper, we use the following definition
\begin{eqnarray}
\xi=\frac{\mu_\chi}{T_\chi},
\end{eqnarray}
to denote the chemical potential of the ADM particle $\chi$ and call $\xi$ chemical potential simply.

\subsection{Effective number of neutrinos}
The energy density of relativistic particles, i.e., radiation components in the early universe, $\rho_{\rm rad}$ is given by
\begin{eqnarray}
\rho_{\rm rad}=\rho_\gamma+\rho_\nu^{\rm std} +\rho_{\rm DR},
\end{eqnarray}
where $\rho_\gamma = (\pi^2/30)g_\gamma T_\gamma^4 $ is the energy density of photons and 
$\rho_\nu^{\rm std}=N_\nu^{\rm std}\rho_\nu= N_\nu^{\rm std}(7/8)(\pi^2/30)g_\nu T_\nu^4$
is the energy density of standard-model massless neutrinos for $N_\nu^{\rm std}$ neutrino families. By the simple estimation in the standard particle cosmology, we have $N_\nu^{\rm std} = 3$. Including the effect of slight reheating of the neutrinos from early $e^+e^-$ annihilation, we obtain $N_\nu^{\rm std} = 3.046$, which is due to the small overlap of neutrino decoupling and $e^+e^-$ annihilation \cite{Mangano2005NPB}. 

The energy density of extra radiation, dark radiation, is generally parametrized through the number of extra effective neutrino species $\Delta N_\nu$ as follows \cite{Lesgourgues2013}
\begin{eqnarray}
\rho_{\rm DR}=\Delta N_\nu \rho_\nu = \Delta N_\nu \frac{7\pi^2}{120}T_\nu^4.
\label{Eq:rho_DR}
\end{eqnarray}
In terms of the following effective number neutrinos
\begin{eqnarray}
N_{\rm eff}=N_\nu^{\rm std}+\Delta N_\nu,
\label{Eq:Neff_def}
\end{eqnarray}
the total energy density of radiation components is estimated to be
\begin{eqnarray}
\rho_{\rm rad}  = \rho_\gamma+N_{\rm eff}\rho_\nu=\left[ 1 + N_{\rm eff}\frac{7}{8}\frac{g_\nu}{g_\gamma} \left(\frac{T_\nu}{T_\gamma}\right)^4 \right] \rho_\gamma. 
\label{Eq:rho_rad}
\end{eqnarray}

As we mentioned in the introduction, if there is an extra light particle as a dark radiation that particle increases the effective number of neutrinos (the direct contribution case) \cite{Sorensen2013PASA, Steigman2013PRD}. On the other hand, although the extra particles are not light enough to be a dark radiation, its annihilation heats other particles via entropy conservation and changes the ratio of $(T_\nu/T_\gamma)$ in Eq.(\ref{Eq:rho_rad}). The extra particle contributes to the total energy density of radiation components $\rho_{\rm rad}$. Consequently, the change of $\rho_{\rm rad}$ reflects to the enhance or the reduce of the effective number of neutrinos $N_{\rm eff}$ via the relation of $\rho_{\rm rad} = \rho_\gamma+N_{\rm eff}\rho_\nu$ even in the absence of dark radiation (the indirect contribution case).

The effective number of neutrino can be probed by its effect on the CMB and the outcome of BBN. The recent observational result on the effective number of neutrinos by the Planck Collaboration is $N_{\rm eff}=3.04 \pm 0.18$ (68\% C.L.) from CMB data \cite{Planck2015arXiv}. On the other hand, we have $N_{\rm eff}=3.71^{+0.47}_{-0.45}$ from BBN data \cite{Steigman2012AHEP}.

\section{Direct contribution}
\label{sec:constraints_direct}

\subsection{Methods}
The method to obtain a constraint on the chemical potential of fermionic DM with the effective number of neutrinos in the case of direct contribution is developed by Boeckel and S.-Bielich \cite{Boeckel2007PRD}. First, we review their method and then we slightly extend their approach to estimate the upper limit on the ADM number asymmetry with the effective number of neutrinos.

Because the energy density of ADM $\rho_{\chi\bar{\chi}}$ cannot excess the energy density of dark radiation $\rho_{DR}$ \cite{Boeckel2007PRD}, we find the following constraint on the energy density of ADM 
\begin{eqnarray}
\frac{g_\chi}{(2\pi)^3}\int E \left[f_\chi(\vec{p},t) + f_{\bar{\chi}}(\vec{p},t) \right]  d^3p
\le  \Delta N_\nu \frac{7\pi^2}{120}T_\nu^4,
\label{Eq:rho_le_rho_DR}
\end{eqnarray}
where $f_i(\vec{p},t)$ is the distribution function of ADM ($i=\chi, \bar{\chi}$). In general, the energy density of fermion increases with its chemical potential. We obtain the upper bound on the chemical potential of ADM, $\mu_\chi^{max}$ as a function of the excess of the effective number of neutrinos $\Delta N_\nu$ at BBN, e.g., $\mu_\chi^{\rm max}=f(\Delta N_\nu)$.

We slightly extend this method to estimate the upper limit on the ADM number asymmetry with the effective number of neutrinos. Because the number density of fermion increases with its chemical potential, the upper bound on the net ADM number density, $n_{\chi\bar{\chi}}=n_{\chi}-n_{\bar{\chi}}$ is a function of the excess of effective number of neutrinos $\Delta N_\nu$ at BBN, e.g., $n_{\chi\bar{\chi}}^{\rm max}=f(\Delta N_\nu)$. We can calculate the upper bound of the ADM number asymmetry at BBN
\begin{eqnarray}
\epsilon_{\rm max}^{\rm BBN}=\frac{n_{\chi\bar{\chi}}^{\rm max}(\Delta N_\nu)}{s},
\end{eqnarray}
where $s$ is the total entropy density $s=\sum_i s_i$. Recall that the number asymmetry is conserved, the upper limit of the ADM number asymmetry at any time $\epsilon_{\rm max}$ , at the epoch of equilibrium $\epsilon^{\rm EQ}_{\rm max}$ and at BBN $\epsilon_{\rm max}^{\rm BBN}$, is the same as $\epsilon_{\rm max}=\epsilon^{\rm EQ}_{\rm max}=\epsilon^{\rm BBN}_{\rm max}$.

\subsection{Constraints}

\begin{figure}[t]
\begin{center}
\includegraphics[width=7.9cm]{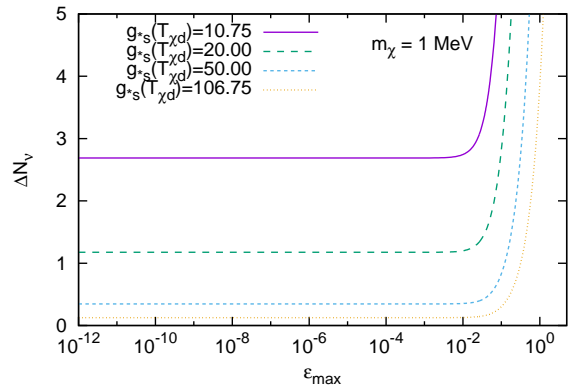}
\caption{The dependence of the extra effective number of neutrinos $\Delta N_\nu$ on the maximum ADM number asymmetry $\epsilon_{\rm max}$.}
\label{fig:direct_fig1}
\end{center}
\end{figure}

For the ultrarelativistic ADM with temperature $T_\chi$, the energy density is obtained as
\begin{eqnarray}
\rho_{\chi\bar{\chi}} = g_\chi T_\chi^4 \left( \frac{7\pi^2}{120}+\frac{1}{4}\xi^2+\frac{1}{8\pi^2}\xi^4\right),
\label{Eq:rho_ADM_rela} 
\end{eqnarray}
and we have
\begin{eqnarray}
g_\chi \left( \frac{7\pi^2}{120}+\frac{1}{4}\xi^2+\frac{1}{8\pi^2}\xi^4\right) \le \Delta N_\nu \frac{7\pi^2}{120}\left(\frac{T_\nu}{T_\chi}\right)^4,
\end{eqnarray}
from the relation of $\rho_{\chi\bar{\chi}} \le \rho_{DR}$. The upper bound of the chemical potential $\mu_\chi$ (more precisely degeneracy parameter $\xi$) of the ultrarelativistic ADM is obtained \cite{Boeckel2007PRD}:
\begin{eqnarray}
\xi_{\rm max}^{\rm BBN}=\sqrt{-\pi^2+\sqrt{\frac{7\pi^4}{15}\left(\frac{\Delta N_\nu }{g_\chi}\left(\frac{T_\nu}{T_\chi}\right)^4+\frac{8}{7} \right)}}.
\label{Eq:xi_max_BBN}
\end{eqnarray}
The number density of ultrarelativistic ADM is estimated as $n_{\chi\bar{\chi}}=(g_\chi T_\chi^3/6\pi^2)(\pi \xi+\xi^3)$ and the upper bound of the ADM number asymmetry is calculated to be
\begin{eqnarray}
\epsilon_{\rm max}=\epsilon_{\rm max}^{\rm BBN} =\frac{g_\chi T_\chi^3}{6\pi^2s}\left[\pi \xi_{\rm max}^{\rm BBN}+(\xi_{\rm max}^{\rm BBN})^3 \right].
\end{eqnarray}

The ratio of the temperature of neutrinos to the temperature of ADM, $T_\nu/T_\chi$, can be estimated by considering conservation of the comoving entropy. At BBN, we have \cite{Boeckel2007PRD}
\begin{eqnarray}
\frac{T_\nu}{T_\chi}=\left(\frac{g_{\ast s}(T_{\chi d})}{g_{\ast s}(T_{\rm BBN})} \right)^{1/3},
\end{eqnarray}
where $T_{\chi d}$ is the ADM decoupling temperature and $T_{\rm BBN}$ is the BBN temperature around $1$ MeV. The effective degrees of freedom at BBN for standard model particles is obtained as $g_{\ast s}(T_{\rm BBN}) = 2 + (7/8)(2\cdot 2 + 2\cdot 3) = 10.75$ for photons, $e^\pm$ and three neutrinos. Including ADM, we have
\begin{eqnarray}
g_{\ast s}(T_{\rm BBN}) = 10.75 + \frac{7}{8}  g_\chi (F_\chi^+ + F_{\bar{\chi}}^+),
\label{Eq:gs_BBN}
\end{eqnarray}
where $F_\chi^\pm$ is given by Eq.(\ref{Eq:Fpm}). We take $F_\chi^+=F_{\bar\chi}^+=1$ and $g_\chi=2$ for the ultra-relativistic Dirac ADM, $g_{\ast s}=14.25$. The effective degrees of freedom at ADM decoupling $g_{\ast s}(T_{\chi d})$ depends on the details of the model of the ADM \cite{Boeckel2007PRD}. If the ADM is decoupled when all standard model particles were present and in equilibrium, we obtain $g_{\ast s}(T_{\chi d})=106.75$. In this section, we treat $g_{\ast s}(T_{\chi d})$ as a free parameter. 

Figure \ref{fig:direct_fig1} shows the dependence of the extra effective number of neutrinos $\Delta N_\nu$ on the maximum ADM number asymmetry $\epsilon_{\rm max}$ with $m_\chi = 1$ MeV. In the case of $\epsilon_{\rm max} \lesssim0.01$, the dependence of $\Delta N_\nu^{\rm max}$ on the $\epsilon_{\rm max}$ is almost negligible, e.g., we obtain the plateau for $\epsilon_{\rm max} \lesssim0.01$ in each curves. The constraint on $\epsilon$ with $\Delta N_\nu$ is obtained:
\begin{eqnarray}
\epsilon_{\rm max}  \simeq 0.01,
\label{Eq:constraint_epsilon_direct}
\end{eqnarray}
for $\Delta N_\nu \lesssim 2.67$ and $ g_{\ast s}(T_{\chi d}) \gtrsim 10.75$.

In the appropriate GeV-scale ADM models, small number asymmetry $\epsilon \simeq 10^{-11}$ is favorable to explain the observed dark matter density $\Omega_{\rm DM}$ (for example, see \cite{Iminniyaz2011JCAP}). For the MeV ADM, the small number asymmetry is still expected (we will consider the relic abundance in Sec. \ref{subsection:Constraints}). The result of $\epsilon_{\rm max}  \simeq 0.01$ is not a strong constraint. The cosmological consequences of $\xi_{\rm max}^{\rm BBN}$ on some problems are important \cite{Boeckel2007PRD}, however, we leave the discussion of the number asymmetry of the ultrarelativistic ADM in the direct contribution case.

Because the nonrelativistic particles role as matter components, the direct contribution of such particles to the energy density of radiation components should be negligible. Consequently, we can hardly probe the chemical potential as well as the asymmetry for nonrelativistic ADM via BBN  \cite{Boeckel2007PRD}. 

\section{Indirect contribution}
\label{sec:constraints_indirect}

\subsection{Methods}
The method to obtain the constraint on the effective number of neutrinos with the symmetric MeV  DM in the indirect contribution is developed in the literature \cite{Ho2013PRD2,Jacques2013PRD,Boeckel2007PRD,Kolb1986PRD,Steigman2013PRD,Nollett2014PRD,Nollett2015PRD,Serpico2004PRD,Ho2013PRD1}. First, we review the useful method for the symmetric MeV DM developed by Steigman \cite{Steigman2013PRD} and then we extend his method to study the ADM in a straightforward way.

{\bf Symmetric DM case (i) $\Gamma_{\chi\gamma, \chi e} \ll \Gamma_{\chi\nu}$: } If the symmetric MeV DM $\chi$ couples more strongly to the neutrinos $\nu$ than to the photons $\gamma$ and electrons $e^-$, e.g. $\Gamma_{\chi\gamma, \chi e} \ll \Gamma_{\chi\nu}$, its late time annihilation heats the neutrinos more than the photons while the annihilation of the $e^\pm$ pairs heats the photons (but not the decoupled neutrinos). The entropy in the comoving volume of photons and $e^\pm$ pairs $S_{\gamma e} = R^3 (s_\gamma + s_e)$ and of the neutrinos and the symmetric MeV DM $S_{\nu \chi} = R^3(s_\nu + s_\chi)$ are conserved individually where $R$ denotes the scale factor \cite{Kolb1990}. The ratio of neutrino to photon temperatures after both of the $e^\pm$ and the symmetric MeV DM are obtained by consideration of entropy conservation as follows:
\begin{equation}
\left( \frac{T_\nu}{T_\gamma}\right)^3 =\frac{g_\gamma[1 + (\tilde{g}_\chi/\tilde{g}_\nu)(\phi_{\chi} / \phi_{\nu})]}{g_\gamma+\tilde{g}_e\phi_{e}}
= \frac{1+\frac{4}{21}\tilde{g}_\chi \phi_{\chi}}{1+\frac{7}{4} \phi_{e}},
\label{Eq:Tnu_Tgamma_chi_nu}
\end{equation}
where $\tilde{g}_i$ denotes effective internal degrees of freedom of the particle $i$; $\tilde{g}_i = (7/8) \cdot 2 \cdot 2 =7/2$ for the Dirac fermions, $\tilde{g}_i = (7/8) \cdot 2 \cdot 1 =7/4$ for the Majorana fermions. Although we consider the only fermionic Dirac type DM, we can also use $\tilde{g}_i=1$ for the scalar bosons and $3$ for the vector bosons with the bosonic distribution function (Eq.(\ref{Eq:f}) with $\Theta_i=-1$). The function of $\phi$ denotes the normalized entropy density 
\begin{eqnarray}
\phi_\alpha (x)=\frac{s^{\rm net}_\alpha (x)}{s^{\rm net}_\alpha  (0)},
\label{Eq:phi}
\end{eqnarray}
where $s^{\rm net}_\alpha (x)$ is the net entropy density of particle species $\alpha$ with vanishing chemical potential \cite{Steigman2013PRD}: 
\begin{equation}
s^{\rm net}_\alpha(x)=s_\alpha (\mu_\alpha=0) + s_{\bar{\alpha}} (\mu_{\bar{\alpha}}=0)
=\sum_{i=\alpha,\bar{\alpha}}\frac{\rho_i+P_i}{T_i}.
\label{Eq:s_rho_p_T}
\end{equation}
For $e^\pm$ pairs, $\phi_{e}$ is evaluated at $x=x_{ed}=m_e/T_{\nu d}$, while for the symmetric DM, $\phi_{\chi}$ is evaluated at $x=x_{\chi d}=m_\chi/T_{\nu d}$ where $T_{\nu d}$ denotes the decoupling temperature of neutrinos. Because the difference between Majorana or Dirac nature of symmetric MeV DM is taken into account by the effective internal degrees of freedom $\tilde{g}_\chi$, we take $g_\chi=1$ in Eq.(\ref{Eq:f}) through our calculations. 

With the appropriate assumptions of $T_{\nu d}=2$ MeV and $\phi_{e}=0.993$ \cite{Steigman2013PRD}, we obtain
\begin{equation}
N_{\rm eff}=3\left[\frac{11}{4}\left( \frac{T_\nu}{T_\gamma}\right)^3 \right]^{4/3}=3.018\left(1+\frac{4\tilde{g}_\chi \phi_{\chi}}{21}\right)^{4/3}.
\label{Eq:Neff_chi_nu}
\end{equation}
If the symmetric MeV DM particle is sufficiently massive, this particle is regarded as a matter component and $\phi_{\chi d} \sim 0$. As a result, the usual result $N_{\rm eff} \simeq 3$ is recovered. On the contrary, for the very light symmetric DM, we obtain $\phi_{\chi} = s_\chi^{\rm net} (x_{\chi d}) / s_\chi^{\rm net} (0) \sim s_\chi^{\rm net} (0) / s_\chi^{\rm net} (0) = 1$ and the excess of the effective number of neutrinos $N_{\rm eff} > 3$ without the dark radiation. The effective number of neutrinos is a function of the mass of symmetric MeV DM.

{\bf Symmetric DM case (ii) $\Gamma_{\chi\gamma, \chi e} \gg\Gamma_{\chi\nu}$: } On the other hand, if the symmetric MeV DM couples more strongly to the electrons and photons than to the neutrinos, $\Gamma_{\chi\gamma, \chi e} \gg\Gamma_{\chi\nu}$, its annihilation heats the electron-photon plasma relative to the neutrino background. By the consideration of entropy conservation, we have
\begin{equation}
\left( \frac{T_\nu}{T_\gamma}\right)^3 =\frac{g_\gamma}{g_\gamma+\tilde{g}_e \phi_{e}+\tilde{g}_\chi \phi_{\chi }}
= \frac{2}{2+\frac{7}{2} \phi_{e}+\tilde{g}_\chi \phi_{\chi}},
\label{Eq:Tnu_Tgamma_chi_gamma_e}
\end{equation}
and 
\begin{equation}
N_{\rm eff}=3\left[\frac{11}{4}\left( \frac{T_\nu}{T_\gamma}\right)^3 \right]^{4/3}
=3\left(\frac{11}{10.95+2\tilde{g}_\chi \phi_{\chi}}\right)^{4/3}. 
\label{Eq:Neff_gamma_e}
\end{equation}
If the symmetric MeV DM particle is sufficiently massive, the usual result $N_{\rm eff} \simeq 3$ is recovered. On the contrary, an annihilation of more light symmetric MeV DM reduces the effective number of neutrinos to be $N_{\rm eff} < 3$.

{\bf ADM case:} We extend the method of calculating $N_{\rm eff}$ with symmetric MeV DM in a simple way. To take care of the chemical potential of the ADM for the entropy calculations, we use the following net entropy density of the ADM 
\begin{eqnarray}
s_\chi^{\rm net}(x,\mu_\chi)=s_\chi + s_{\bar{\chi}}=\sum_{i=\chi,\bar{\chi}}\frac{\rho_i+P_i-\mu_i n_i}{T_i},
\end{eqnarray}
instead of Eq.(\ref{Eq:s_rho_p_T}). The normalized entropy density of the ADM to be a function of not only $x_{\chi d}$ but also $\mu_\chi$ is given as follows:
\begin{eqnarray}
\phi_\chi (x_{\chi d}, \mu_\chi)=\frac{s_\chi^{\rm net}(x_{\chi d}, \mu_\chi)}{s_\chi^{\rm net}(0,\mu_\chi)}.
\end{eqnarray}
Other quantities in Eq.(\ref{Eq:Tnu_Tgamma_chi_nu}) and Eq.(\ref{Eq:Tnu_Tgamma_chi_gamma_e}), such as $\tilde{g}_\chi, \phi_e$, remain the same. 

The effective number of neutrinos with the MeV ADM is estimated in the same form of Eq.(\ref{Eq:Neff_chi_nu}) and Eq.(\ref{Eq:Neff_gamma_e}). If the MeV ADM $\chi$ couples more strongly to the neutrinos $\nu$ than to the photons $\gamma$ and electrons $e^-$ ($\Gamma_{\chi\gamma, \chi e} \ll \Gamma_{\chi\nu}$), we obtain
\begin{eqnarray}
N_{\rm eff}&=&3.018\left(1+\frac{4\tilde{g}_\chi \phi_\chi (x_{\chi d}, \mu_\chi)}{21}\right)^{4/3},
\label{Eq:ADM_Neff_chi_nu}
\end{eqnarray}
while if the MeV ADM couples more strongly to the electrons and photons than to the neutrinos ($\Gamma_{\chi\gamma, \chi e} \gg\Gamma_{\chi\nu}$), we have
\begin{eqnarray}
N_{\rm eff}
&=&3\left(\frac{11}{10.95+2\tilde{g}_\chi \phi_\chi (x_{\chi d}, \mu_\chi)}\right)^{4/3}.
\label{Eq:ADM_Neff_gamma_e}
\end{eqnarray}

These derivations of the effect of light ADM on the effective number of ultrarelativistic species  at the epoch of BBN, Eqs.(\ref{Eq:ADM_Neff_chi_nu}) and (\ref{Eq:ADM_Neff_gamma_e}), are the main new findings in this paper.

In the following numerical calculations, we vary the mass of the MeV ADM in the range of $m_\chi = 0.1 - 100$ MeV. In this case, both of the ultrarelativistic limit and the nonrelativistic limit are inappropriate. We estimate the thermodynamic quantities such as the number density from the first principle (most fundamental formulae) in Eq.(\ref{Eq:n}), Eq.(\ref{Eq:rho}), Eq.(\ref{Eq:P}) and Eq.(\ref{Eq:s}) numerically by the  Gauss-Laguerre integration method:
\begin{eqnarray}
\int_0^\infty x^\alpha e^{-x} F(x) dx = \sum_{j=1}^N w_j F(x_j),
\label{Eq:GaussLaguerre}
\end{eqnarray}
where $F(x)$ and $w_j$ denote a function of $x$ and ``Gauss-Laguerre weights," respectively \cite{Press1992}. In our calculation, $\alpha=0$ and $N=50$ are chosen. In order to apply the formula in Eq.(\ref{Eq:GaussLaguerre}) to calculate the number density $n_i$, we estimate the following integral \cite{Schwarz2009JCAP,Iizuka2015MPLA}
\begin{eqnarray}
n_i=\frac{1}{2\pi^2}\int_{0}^\infty T_i(T_ix+m_i)[(T_ix+m_i)^2-m_i^2]^{1/2} f_i dx,\nonumber \\
\label{Eq:gauss_laguerre_n}
\end{eqnarray}
instead of Eq.(\ref{Eq:n}), where $x=(E_i-m_i)/T_i$ and the distribution function in Eq.(\ref{Eq:f}) is replaced by
\begin{eqnarray}
f_i=\frac{g_i}{e^{(T_ix+m_i-\mu_i)/T_i}+\Theta_i}.
\label{Eq:gauss_laguerre_f}
\end{eqnarray}
Similarly, the energy density $\rho_i$ and the pressure $P_i$ are calculated as
\begin{eqnarray}
\rho_i&=&\frac{1}{2\pi^2}\int_{0}^\infty T_i(T_ix+m_i)^2[(T_ix+m_i)^2-m_i^2]^{1/2} f_i dx, \nonumber \\
P_i&=&\frac{1}{6\pi^2}\int_{0}^\infty T_i[(T_ix+m_i)^2-m_i^2]^{3/2} f_i dx,\label{Eq:gauss_laguerre_rho_P}
\end{eqnarray}
instead of Eq.(\ref{Eq:rho}) and Eq.(\ref{Eq:P}), respectively.

Before we show the results from the numerical calculations, we would like to comment on the role of asymmetry at BBN. The effect of the annihilating thermal relic particles to heat either the photons or the light particles. The bulk of this heat is created when the temperature is within a factor of a few of the relic particle mass $T \sim m_\chi$. At these temperatures the number asymmetry between particles and antiparticles may be small without large chemical potential. At these temperatures the number of particle and antiparticle pairs could be much larger than the relic number when annihilations have ceased, at a temperature much lower than the particle mass. Thus, the effect of the number asymmetry of unobserved MeV scale particles is not strong at BBN (except the very light and degenerate (asymmetric) neutrinos. For example, see Refs.\cite{Ichimasa2014PRD,Ichikawa2003PLB}).

Although the case of asymmetric dark matter is not significantly different from the case of symmetric dark matter without large chemical potential of ADM, we would like to announce that the explicit considerations of the relation between the chemical potential of ADM and effective number of neutrinos at BBN are first shown in this paper. Some numerical results are expected without actual calculations, however, the actual calculations are important complement to any previously reported results in the literature. 

\subsection{Constraints}
\label{subsection:Constraints}
\begin{figure}[t]
\begin{center}
\includegraphics[width=7.9cm]{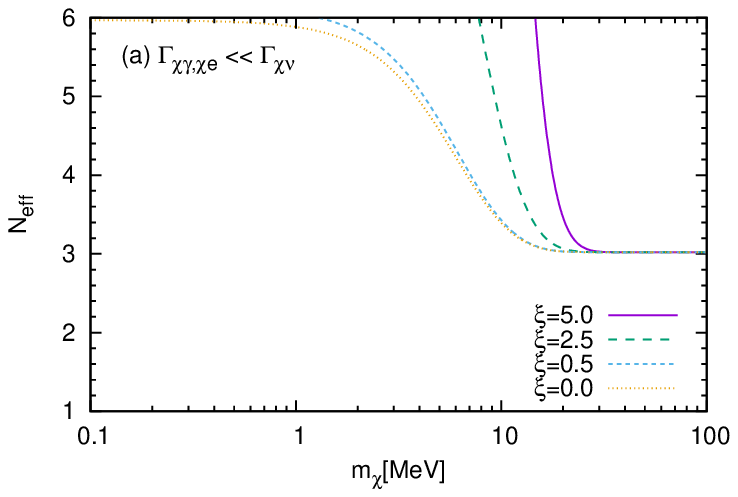}
\includegraphics[width=7.9cm]{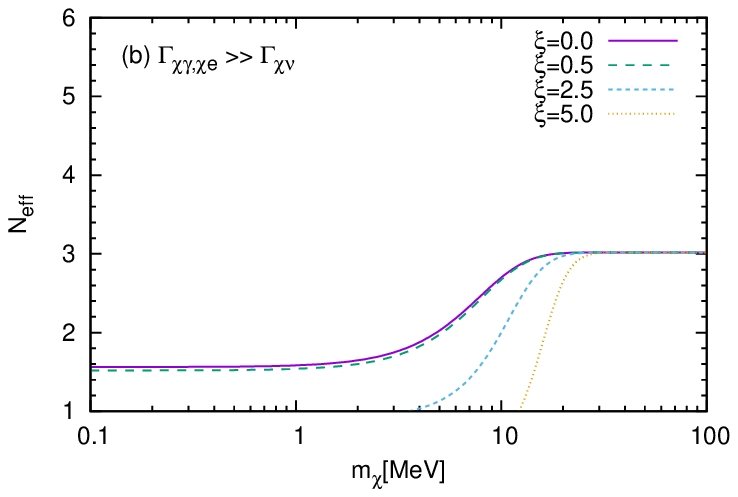}
\caption{The dependence of the effective number of neutrinos $N_{\rm eff}$ on the ADM mass $m_\chi$ MeV with the various chemical potential $\xi$ at BBN. The figure (a) shows the dependence in the case of $\Gamma_{\chi\gamma, \chi e} \ll \Gamma_{\chi\nu}$ while the figure (b) shows the dependence in the case of $\Gamma_{\chi\gamma, \chi e} \gg \Gamma_{\chi\nu}$. }
\label{fig:indirect_fig1}
\end{center}
\end{figure}

{\bf 1. Lower mass limits:} Figure \ref{fig:indirect_fig1} shows the dependence of the effective number of neutrinos $N_{\rm eff}$ on the mass of the ADM $m_\chi$ with the various chemical potentials $\xi$ at BBN ($T=2$ MeV). The figure (a) shows the dependence in the case of $\Gamma_{\chi\gamma, \chi e} \ll \Gamma_{\chi\nu}$ while the figure (b) shows the dependence in the case of $\Gamma_{\chi\gamma, \chi e} \gg \Gamma_{\chi\nu}$. 

The value of $N_{eff}$ with the vanishing chemical potential ($\xi=0$) is compatible with the previously reported results for the symmetric MeV DM case \cite{Steigman2013PRD,Nollett2015PRD,Boehm2012JCAP,Boehm2013JCAP}. For example, B\oe hm, et al. obtained the lower bound on the mass of Dirac fermion DM from the CMB data (and BBN considerations) at 95\% C.L. as $m_\chi > 7.3$ MeV for $N_{eff} =3.30^{+0.54}_{-0.51}=2.79-3.84$ in the case of $\Gamma_{\chi\gamma, \chi e} \ll \Gamma_{\chi\nu}$ \cite{Boehm2013JCAP}. We obtain the same lower limit of $m_\chi$ for $N_{eff}=3.86$ from Fig.\ref{fig:indirect_fig1}. We note that the value of $N_{eff}$ at CMB and at BBN may be different, e.g., a MeV scale particle with the vanishing chemical potential (mass $\lesssim 10$ MeV) may lead to the value of $N_{eff}$ at CMB formation which is lager than at BBN \cite{Boehm2012JCAP}. In this paper, we take $N_{eff}$ to be defined at BBN.

The similar figure of Fig.\ref{fig:indirect_fig1} has already reported by Nollett and Steigman for symmetric DM \cite{Nollett2014PRD,Nollett2015PRD} . The curves with nonvanishing chemical potential in Fig.\ref{fig:indirect_fig1} are newly obtained in our study. The dependence of the chemical potentials of ADM on the effective number of neutrinos is shown explicitly for the first time in this paper. Compere with the symmetric DM case, the effective number of neutrinos $N_{\rm eff}$ increases with the increasing asymmetry (chemical potential $\xi$) and with the decreasing mass $m_\chi$ if ADM particle mainly interacts to neutrinos ($\Gamma_{\chi\gamma, \chi e} \ll \Gamma_{\chi\nu}$). On the contrary,  $N_{\rm eff}$ decreases with the increasing $\xi$ and with the decreasing $m_\chi$ if ADM particle mainly interacts to photons and electrons ($\Gamma_{\chi\gamma, \chi e} \gg \Gamma_{\chi\nu}$). Also, in the Nollett and Steigman papers for symmetric DM \cite{Nollett2014PRD,Nollett2015PRD}, $N_{\rm eff}$ depends on the nature of the quantum statistics of thermal relic (i.e., fermion or boson). We can take the effective internal degrees of freedom as $\tilde{g}_\chi = 1, 7/4, 2, 7/2$ for a real scalar, Majorana fermion, complex scalar and Dirac fermion in Eqs(\ref{Eq:ADM_Neff_chi_nu}) and (\ref{Eq:ADM_Neff_gamma_e}). Thus, $N_{\rm eff}$ depends on the difference of the nature of the thermal relic not only in the symmetric DM case but also in the ADM case. 

The lower mass limit is obtained with the upper bound or lower bound on the effective number of neutrinos. For example, we obtain $m_\chi \gtrsim m_\chi^{\rm min} =18.1$ if $N_{\rm eff}^{\rm max}=3.0$ in the figure (a), while $m_\chi \gtrsim m_\chi^{\rm min} =18.3$ if $N_{\rm eff}^{\rm min}=3.0$ in the figure (b). Thus, the sets of $\{(\xi, m_\chi^{\rm min})\}$ for the fixed $N_{\rm eff}^{\rm max}$ or $N_{\rm eff}^{\rm min}$ is obtained. The lower bound on the ADM mass is the case when the asymmetry and chemical potential are small. The bound gets stronger as asymmetry $\epsilon$ and chemical potential $\xi$ grow. It shows the smooth transition from asymmetric to the symmetric WIMP limit.

\begin{figure}[t]
\begin{center}
\includegraphics[width=7.9cm]{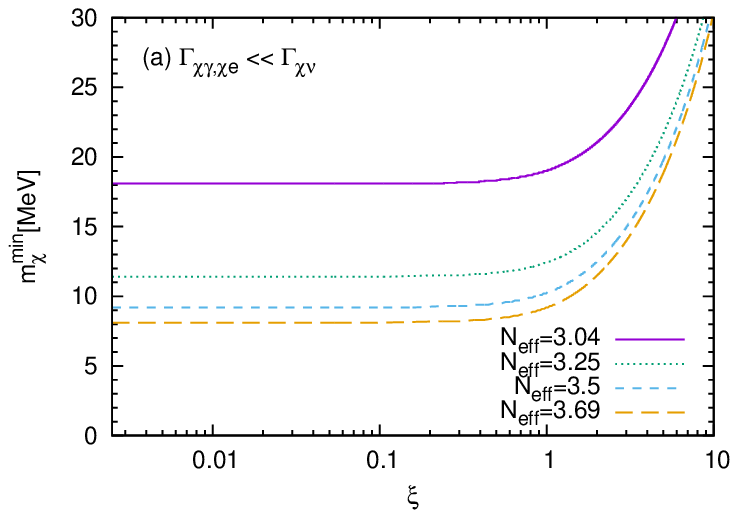}
\includegraphics[width=7.9cm]{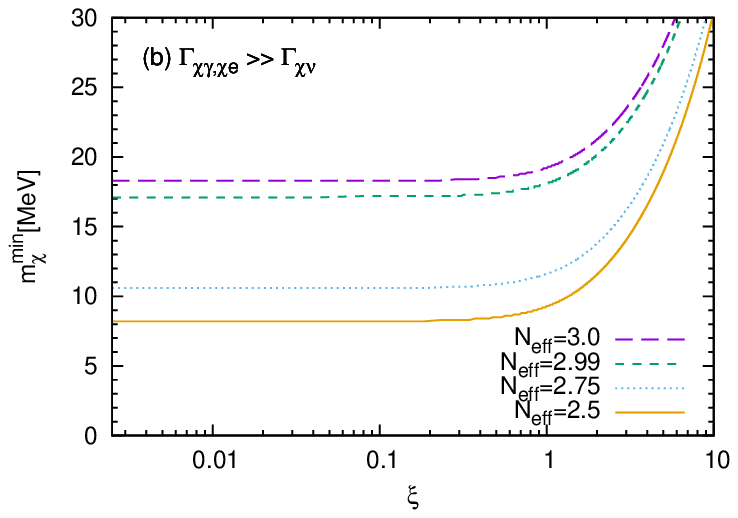}
\caption{The dependence of the lower mass limit of the ADM $m_\chi^{\rm min}$ on the chemical potential $\xi$ with the various effective number of neutrinos $N_{\rm eff}$ at BBN ($T=2$ MeV). The figure (a) shows the $\Gamma_{\chi\gamma, \chi e} \ll \Gamma_{\chi\nu}$ case while the figure (b) shows the $\Gamma_{\chi\gamma, \chi e} \gg \Gamma_{\chi\nu}$ case.}
\label{fig:indirect_fig2}
\end{center}
\end{figure}

The sets of $\{(\xi, m_\chi^{\rm min})\}$ is shown in Fig.\ref{fig:indirect_fig2}. This figure shows the dependence of the lower mass limit of the ADM $m_\chi^{\rm min}$ on the chemical potential $\xi$ with the various bound of the effective number of neutrinos $N_{\rm eff}$ at BBN. The figure (a) shows the $\Gamma_{\chi\gamma, \chi e} \ll \Gamma_{\chi\nu}$ case while the figure (b) shows the $\Gamma_{\chi\gamma, \chi e} \gg \Gamma_{\chi\nu}$ case. The ADM mass $m_\chi^{\rm min}$ increases with the decreasing upper bound of the effective number of neutrinos $N_{\rm eff}^{\rm max}$ in the $\Gamma_{\chi\gamma, \chi e} \ll \Gamma_{\chi\nu}$ case. On the contrary,  $m_\chi^{\rm min}$ increases with the increasing the lower bound of $N_{\rm eff}^{\rm min}$ in the $\Gamma_{\chi\gamma, \chi e} \gg \Gamma_{\chi\nu}$ case. Moreover $m_\chi^{\rm min}$ increases with the increasing chemical potential $\xi$ in both cases.

\begin{figure}[t]
\begin{center}
\includegraphics[width=7.9cm]{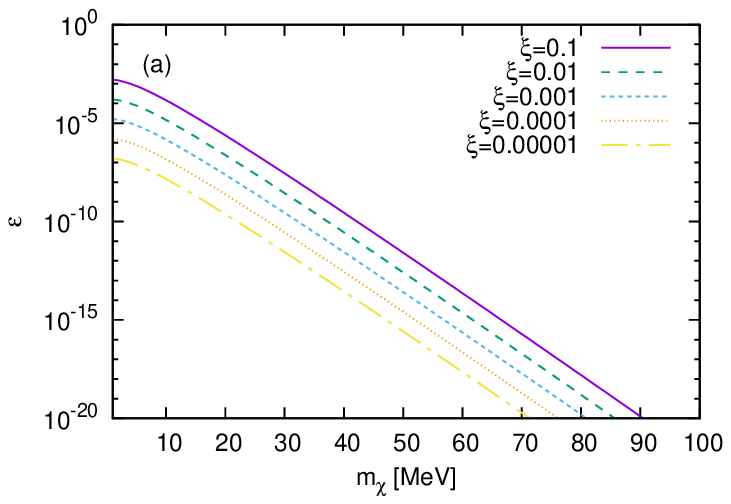}
\includegraphics[width=7.9cm]{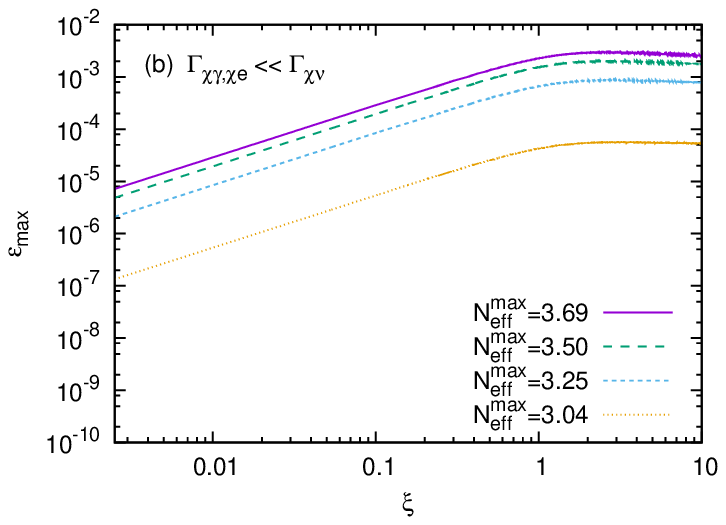}
\includegraphics[width=7.9cm]{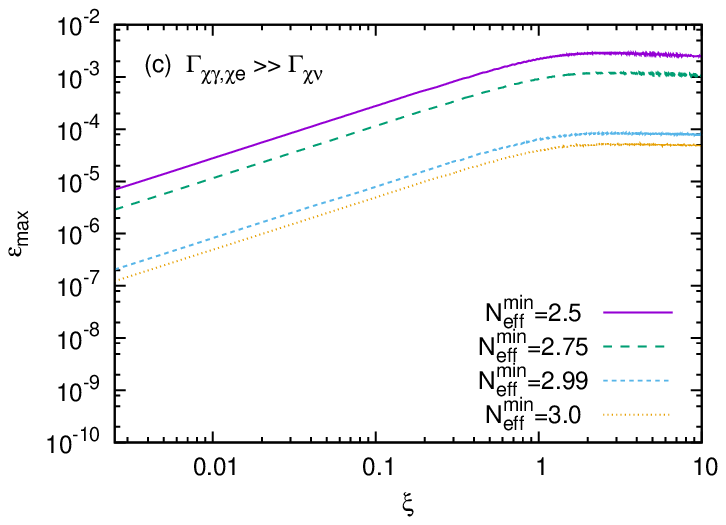}
\caption{ADM number asymmetry. Figure (a): the dependence of the ADM number asymmetry $\epsilon_{\rm}$ on the mass $m_\chi$ with fixed chemical potential $\xi$. Figure (b) and (c): the dependence of the upper limit of the ADM number asymmetry $\epsilon_{\rm max}$ on the chemical potential $\xi$ with the various bound of the effective number of neutrinos $N_{\rm eff}^{\rm max}$ or $N_{\rm eff}^{\rm min}$ at BBN. The figure (b) shows the $\Gamma_{\chi\gamma, \chi e} \ll \Gamma_{\chi\nu}$ case while the figure (c) shows the $\Gamma_{\chi\gamma, \chi e} \gg \Gamma_{\chi\nu}$ case.}
\label{fig:indirect_fig3}
\end{center}
\end{figure}

From Fig.\ref{fig:indirect_fig2}, the following constraint on $m_\chi^{\rm min}$ with $N_{\rm eff}$ is obtained:
\begin{eqnarray}
m_\chi^{\rm min} \simeq
\begin{cases}
8.1 - 18.1 \ {\rm MeV} & (N_{\rm eff}^{\max}=3.0-3.7) \\
8.2 - 18.3 \ {\rm MeV} & (N_{\rm eff}^{\min}=2.5-3.0) 
\end{cases}
\label{Eq:constraint_m_chi_min_indirect}
\end{eqnarray}
for $\Gamma_{\chi\gamma, \chi e} \ll \Gamma_{\chi\nu}$ (upper) and $\Gamma_{\chi\gamma, \chi e} \gg \Gamma_{\chi\nu}$ (lower) cases, more concretely,
\begin{eqnarray}
m_\chi^{\rm min} \simeq
\begin{cases}
18.1 \ {\rm MeV} & (\Gamma_{\chi\gamma, \chi e} \ll \Gamma_{\chi\nu},N_{\rm eff}^{\rm max} = 3.0) \\
18.3 \ {\rm MeV} & (\Gamma_{\chi\gamma, \chi e} \gg \Gamma_{\chi\nu},N_{\rm eff}^{\rm min} = 3.0) 
\end{cases}
\label{Eq:constraint_m_chi_min_indirect_neff3}
\end{eqnarray}
with the standard value of the effective number of neutrinos.

{\bf 2. Maximum number asymmetry:} Once the sets of $\{(\xi, m_\chi^{\rm min})\}$ is obtained, we can estimate the ADM number asymmetry at $(\xi, m_\chi^{\rm min})$:
\begin{eqnarray}
\epsilon (\xi,m_\chi^{\rm min}) = \frac{1}{s}\left[n_\chi(\xi,m_\chi^{\rm min})-n_{\bar{\chi}}(\xi,m_\chi^{\rm min})\right].
\end{eqnarray}
The total entropy density $s$ is calculated as $s = s_{\rm SM} ^{\rm net}+ s^{\rm net}_{\chi}$ where $s^{\rm net}_{\rm SM}\simeq s^{\rm net}_\gamma + s^{\rm net}_e + 3 s^{\rm net}_\nu$ denotes the sum of the net entropy densities of the standard model particles at BBN and $s^{\rm net}_{\chi}$ denotes the net entropy density of the ADM $\chi$. In the numerical calculation of the entropy density [see Eq.(\ref{Eq:s}) with Eq.(\ref{Eq:gauss_laguerre_n}) and Eq.(\ref{Eq:gauss_laguerre_rho_P})], the masses of all particles except neutrinos are taken from the particle data group \cite{PDG2014}. Although, the recent results of the Planck experiment give us a constraint on the sum of the light neutrino masses as $\sum m_{\nu} < 0.17$ eV  (95\% C.L.) \cite{Planck2015arXiv}, we assume that all neutrino masses are the same as $m_\nu=m_{\nu_e}=m_{\nu_\mu}=m_{\nu_\tau}=1$ eV in our calculation for the sake of simplicity. The unknown tiny masses of the neutrinos are irrelevant in our study. Moreover, we neglect the chemical potential of all standard model particles. 

Because the ADM number asymmetry $\epsilon$ decreases with the increasing mass $m_\chi$ (we will see below), we obtain the upper limit of the ADM number asymmetry 
\begin{eqnarray}
\epsilon_{\rm max} =\epsilon (\xi,m_\chi^{\rm min}),
\end{eqnarray}
and the set of $\{(\xi, \epsilon_{\rm max})\}$ for the fixed $N_{\rm eff}^{\rm max}$ or  $N_{\rm eff}^{\rm min}$. 

Figure \ref{fig:indirect_fig3} shows the ADM number asymmetry. Figure (a) describes the dependence of the ADM number asymmetry $\epsilon_{\rm}$ on the mass $m_\chi$ with fixed chemical potential $\xi$. Figures (b) and (c) describe the dependence of the upper limit of the ADM number asymmetry $\epsilon_{\rm max}$ on the chemical potential $\xi$ with the various bound of the effective number of neutrinos $N_{\rm eff}^{\rm max}$ or $N_{\rm eff}^{\rm min}$ at BBN. The figure (b) shows the $\Gamma_{\chi\gamma, \chi e} \ll \Gamma_{\chi\nu}$ case while the figure (c) shows the $\Gamma_{\chi\gamma, \chi e} \gg \Gamma_{\chi\nu}$ case. In the figure (a), we see that the ADM number asymmetry $\epsilon$ decreases with the increasing mass $m_\chi$. The figure (b) shows that the upper bound of the ADM number asymmetry $\epsilon_{\rm max}$ increases with the increasing effective number of neutrinos $N_{\rm eff}^{\rm max}$ in the $\Gamma_{\chi\gamma, \chi e} \ll \Gamma_{\chi\nu}$ case. On the contrary,  the figure (c) shows that $\epsilon_{\rm max}$ increases with the decreasing $N_{\rm eff}^{\rm min}$ in the $\Gamma_{\chi\gamma, \chi e} \gg \Gamma_{\chi\nu}$ case. 

From Fig.\ref{fig:indirect_fig3}, the following constraint on $\epsilon_{\rm max}$ with $N_{\rm eff}$ is obtained:
\begin{eqnarray}
&&\epsilon_{\rm max} \simeq \nonumber \\
&&\quad
\begin{cases}
1.35 \times 10^{-7} - 7.20 \times 10^{-6} & (N_{\rm eff}^{\max}=3.0-3.7) \\
1.23 \times 10^{-7} - 6.95 \times 10^{-6} & (N_{\rm eff}^{\min}=2.5-3.0) 
\end{cases}
\nonumber \\
&&\quad {\rm for} \quad \xi\gtrsim 0.0025,
\label{Eq:constraint_epsilon_max_indirect}
\end{eqnarray}
for $\Gamma_{\chi\gamma, \chi e} \ll \Gamma_{\chi\nu}$ (upper) and $\Gamma_{\chi\gamma, \chi e} \gg \Gamma_{\chi\nu}$ (lower) cases, more concretely,
\begin{eqnarray}
\epsilon_{\rm max} \simeq
\begin{cases}
1.35 \times 10^{-7}  & (\Gamma_{\chi\gamma, \chi e} \ll \Gamma_{\chi\nu},N_{\rm eff}^{\rm max} = 3.0) \\
1.23 \times 10^{-7}   & (\Gamma_{\chi\gamma, \chi e} \gg \Gamma_{\chi\nu},N_{\rm eff}^{\rm min} = 3.0)
\end{cases}
\label{Eq:constraint_epsilon_max_indirect_neff3}
\end{eqnarray}
with the standard value of the effective number of neutrinos.

We note that, as shown in figure (b) and figure (c), $\epsilon_{\rm max}$ increases with the increasing chemical potential $\xi$. It is naturally expected because the origin of the ADM number asymmetry is the nonvanishing chemical potentials of $\chi$ and $\bar{\chi}$. Although there is no indication of the upper limit of the ADM number asymmetry $\epsilon_{\rm max}$ for $\xi=0$ in Fig.\ref{fig:indirect_fig3}, $\epsilon_{\rm max}=0$ is obtained correctly for $\xi=0$.

As we show below, in consideration of the relic abundance of the MeV ADM, the obtained upper limit on the ADM number asymmetry in Eq.(\ref{Eq:constraint_epsilon_max_indirect}) and Eq.(\ref{Eq:constraint_epsilon_max_indirect_neff3}) is not a strong constraint, if the origin of the observed 511 keV gamma-ray is the annihilating light MeV ADM.

{\bf 3. Relic abundance:} The relic abundance of the ADM $\Omega_{\rm ADM} h^2$ depends on not only the ADM number asymmetry $\epsilon$ but also the thermally averaged annihilation cross section $\langle \sigma_{\chi\bar\chi}v\rangle$ [see Eq.(\ref{Eq:BoltzmannEq_final})]. The averaged cross section is obtained by \cite{Gondolo1991NPB}
\begin{eqnarray}
\langle \sigma_{\chi\bar\chi}v\rangle &=& \frac{1}{8m_\chi^4 T_\chi K_2^2(x)} \\
&& \times \int_{4m_\chi^2}^\infty \sigma_{\chi\bar\chi}(s)(s-4m_\chi^2)\sqrt{s}K_1(\sqrt{s}/T_\chi)ds, \nonumber
\end{eqnarray}
where  $\sigma_{\chi\bar\chi}$, $K_i$, $s$ and $x$ denote the annihilation cross section, the modified Bessel function of order $i$, one of the Mandelstam variable (not an entropy density) and $x=m_\chi/T_\chi$, respectively. Practically, the averaged cross section can be expanded in power of the relative velocity of incoming particles $v$. The standard approximation of the averaged cross section is \cite{Scherrer1986PRD,Ellwanger2012JCAP}:
\begin{eqnarray}
\langle \sigma_{\chi\bar\chi}v\rangle \simeq a + bx^{-1} + \mathcal{O}(x^{-2}).
\end{eqnarray}
If the s-wave annihilation is dominant, we can take $a \neq 0$ and $b=0$. On the other hand, if the p-wave annihilation is dominant, we can put $a=0$ and $b\neq 0$. In this study, the coefficients $a$ and $b$ are taken as free parameters to perform a model independent analysis.   

\begin{figure}[t]
\begin{center}
\includegraphics[width=7.8cm]{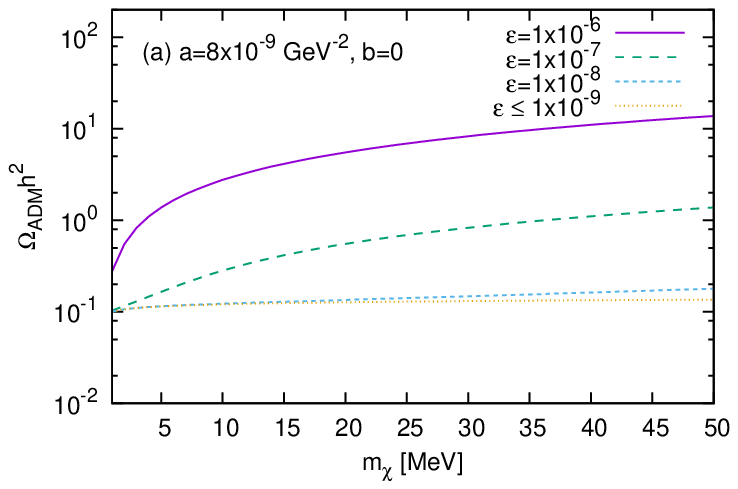}
\includegraphics[width=7.8cm]{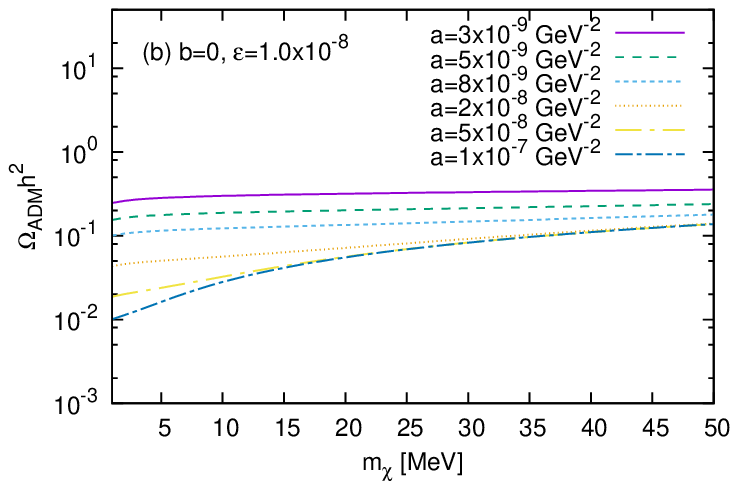}
\includegraphics[width=7.8cm]{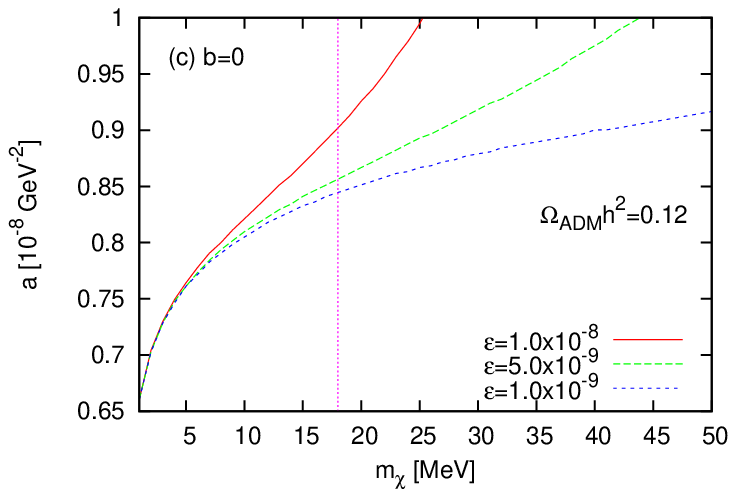}
\caption{The dependence of the relic abundance of the MeV ADM $\Omega_{\rm ADM} h^2$ on the mass $m_\chi$ in the s-wave dominant case. The figure (a) shows the dependence with the various number asymmetry $\epsilon$, while the figure (b) shows the dependence with the various cross section $\langle \sigma_{\chi\bar\chi}v\rangle \simeq a + bx^{-1}$. The figure (c) shows the value of parameter ``$a$" required to satisfy the observed relic abundance of DM, $\Omega_{\rm ADM}h^2 = 0.12$ and lower mass limit for the $N_{eff}=3.0$. The vertical line in figure (c) shows the lower mass limit.}
\label{fig:indirect_fig4}
\end{center}
\end{figure}
\begin{figure}[t]
\begin{center}
\includegraphics[width=7.8cm]{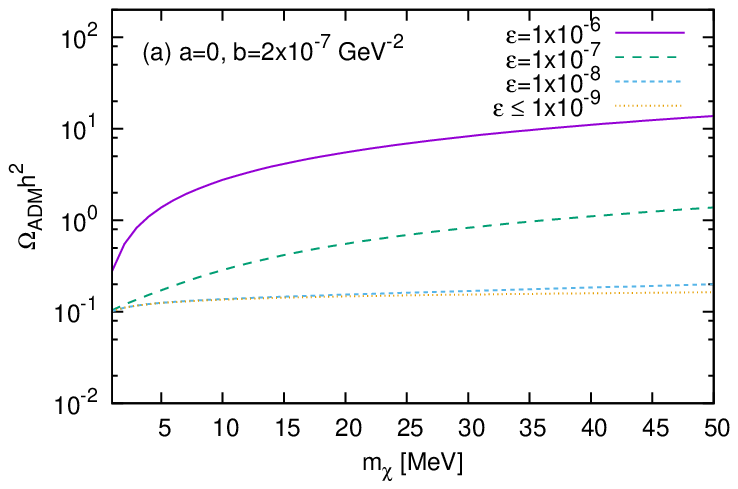}
\includegraphics[width=7.8cm]{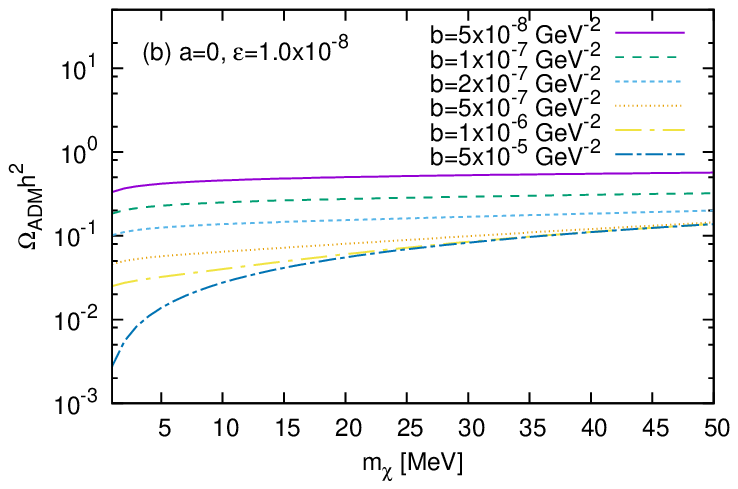}
\includegraphics[width=7.8cm]{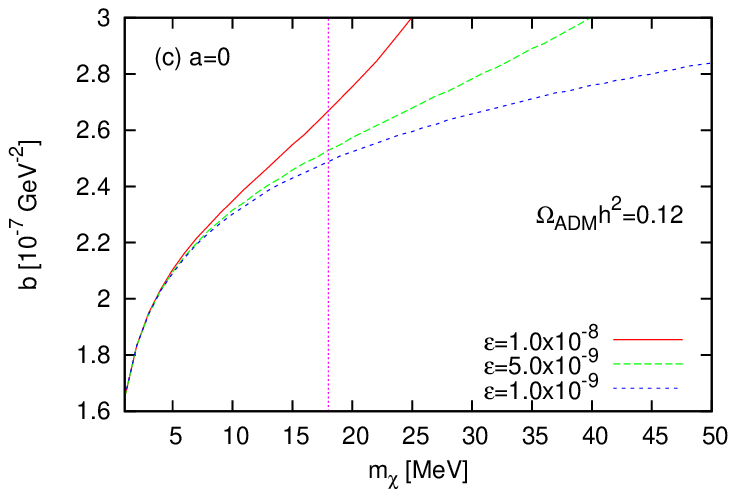}
\caption{Same as Fig.\ref{fig:indirect_fig4} but p-wave dominant case.}
\label{fig:indirect_fig5}
\end{center}
\end{figure}

Figures \ref{fig:indirect_fig4} and \ref{fig:indirect_fig5} show the dependence of the relic abundance of the MeV ADM $\Omega_{\rm ADM} h^2$ on the mass $m_\chi$ in the s-wave dominant case (Fig.\ref{fig:indirect_fig4}) and in the p-wave dominant case (Fig.\ref{fig:indirect_fig5}), respectively. In Fig.\ref{fig:indirect_fig4}, the figure (a) shows the dependence with the various number asymmetry $\epsilon$, while the figure (b) shows the dependence with the various cross section $\langle \sigma_{\chi\bar\chi}v\rangle \simeq a + bx^{-1}$. The figure (c) shows the value of parameter ``$a$" required to satisfy the observed relic abundance of DM, $\Omega_{\rm ADM}h^2 = 0.12$ and lower mass limit for the $N_{eff}=3.0$. The vertical line in figure (c) shows the lower mass limit. The relic abundance of the MeV ADM $\Omega_{\rm ADM} h^2$ decreases with the decreasing ADM number asymmetry $\epsilon$ as shown in the figure (a). The mass dependence on the $\Omega_{\rm ADM} h^2$ is significant for the very light $m_\chi \lesssim 5$ MeV and the highly asymmetric $\epsilon \gtrsim 1\times 10^{-7}$ case. In other words, we can ignore the mass dependence on the relic abundance for the case of $\epsilon \lesssim 1\times 10^{-8}$. Indeed, for $\epsilon \lesssim 1\times 10^{-8}$, the s-wave dominant cross section is almost degenerate for $\Omega_{\rm ADM}h^2 = 0.12$ as shown in figure (c). The behavior of the relic abundance in the p-wave dominant case (Fig.\ref{fig:indirect_fig5}) is similar to the s-wave dominant case (Fig.\ref{fig:indirect_fig4}).

The cosmological constraints on the annihilation cross section of the symmetric DM are extensively studied in the literature, such as the constraints from the galactic 511 keV gamma-ray line  \cite{Boehm2004PRL,Fayet2004PRD,Beacom2005PRL,Boehm2006PLB,Jacoby2007JHEP,Pospelov2007PLB,Finkbeiner2007PRD,Huh2008PRD,Lingenfelter2009PRL,Vincednt2012JCAP,Huang2013PRD}, from the galactic 3.5 keV x-ray line \cite{Graham2015PRD,Patra2015PRD,Arcadi2015JCAP}, from other cosmic rays  \cite{Cline2012PLB,Masina2012JCAP,Giesen2015ArXiv,McDermott2015PDU,Regis2015PRL,Sameth2015PRL}, from the CMB \cite{McDonald2000PRD,Ali-Haimound2015PRL,DAgnolo2015PRL,Steigman2015PRD} and from the BBN consideration \cite{Frieman1990PRD}. 

For the symmetric MeV DM, the significant constraints on the annihilation cross section are obtained from the galactic 511 keV gamma-ray line observation. The light particles $\chi$ ($m_\chi \sim 1 - 100$ MeV) annihilating into $e^+e^-$ pairs in the galactic bulge may be the source of the observed gamma-ray \cite{Boehm2004PRL}. According to this scenario, if the s-wave annihilation is dominant, $m_\chi \sim 100$ MeV is required to obtain the observed cosmic ray flux without the final state radiation/bremsstrahlung, the so-called internal bremsstrahlung (IB), effects. Including IB effects, we may expect that the upper bound of the symmetric dark matter particles is around 3 MeV \cite{BeacomPRL2006} or 7.5 MeV \cite{Sizun2006PRD}. On the contrary, if the cross section is p-wave dominant, the lighter DM ($m_\chi \sim 1$ MeV) is possible for $b \sim 10-200$ pb ($b \sim 2.6\times 10^{-8} - 5.1\times 10^{-7}$ GeV$^{-2}$) \cite{Boehm2004PRL,Jacoby2007JHEP}. 

If the s-wave annihilation is dominant, the upper bound of symmetric dark matter $m_\chi \lesssim 3$ or $7.5$ MeV is at almost margin of our result in Eq.(\ref{Eq:constraint_m_chi_min_indirect}) for $N_{eff}=2.5-3.7$. Moreover, for $N_{eff}=3.0$, $m_\chi \lesssim 3$ or $7.5$ MeV is not consistent with our result $m_\chi^{\rm min} \sim 18$ MeV in Eq.(\ref{Eq:constraint_m_chi_min_indirect_neff3}). Although, we may expect that the constraint on the symmetric MeV DM is also appropriate for the MeV ADM approximately, there is a little study for the cosmological constraints on the annihilation cross section of the ADM \cite{Boehm2004NPB}. The upper mass limit of ADM may be a few MeV, however, we take a conservative upper limit of the mass of ADM as a few $10$ MeV in this paper.

Figure \ref{fig:indirect_fig4} shows that the s-wave dominant annihilation with $a\simeq 8\times 10^{-9}$ GeV$^{-2}$, $b=0$ and $\epsilon \simeq 1.0\times 10^{-8}$ is consistent with the observed energy density of the cold dark matter component $\Omega_{\rm DM}h^2\simeq 0.1$. For the p-wave dominant annihilation, $a=0$, $b\simeq 3\times 10^{-7}$ GeV$^{-2}$ and $\epsilon \simeq 1.0\times 10^{-8}$ is consistent with data as shown in Fig.\ref{fig:indirect_fig5}. If the origin of the observed 511 keV gamma-ray is the annihilating MeV ADM and/or all of the cold dark matter is made of the MeV ADM $\Omega_{\rm ADM}h^2 = \Omega_{\rm DM}h^2$, the ADM number asymmetry $\epsilon$ should be less than the obtained upper limit $\epsilon_{\rm max}$ in Eq.(\ref{Eq:constraint_epsilon_max_indirect}) and Eq.(\ref{Eq:constraint_epsilon_max_indirect_neff3}) at least under $10^{-1}$. If the effective number of neutrinos is just the standard value $N_{\rm eff}=3.0$, the ADM number asymmetry should be $\epsilon \simeq \epsilon_{\rm max} \times 10^{-3}$. As a consequence, the upper limit on the ADM number asymmetry in Eq.(\ref{Eq:constraint_epsilon_max_indirect}) and Eq.(\ref{Eq:constraint_epsilon_max_indirect_neff3}) is not strong constraint.

We comment that the constraints in Eq.(\ref{Eq:constraint_epsilon_max_indirect}) and Eq.(\ref{Eq:constraint_epsilon_max_indirect_neff3}) may be more strict in the combination of the following cases: (a) if the chemical potential of the ADM is tiny $\xi\ll 0.0025$, (b) if the origin of the observed 511 keV gamma-ray is not the annihilating MeV ADM and (c)  the MeV ADM is a part of the cold dark matter $\Omega_{\rm ADM}h^2 < \Omega_{\rm DM}h^2$.

\section{Summary}
\label{sec:summary}
We have extended the known two methods by Boeckel and S.-Bielich \cite{Boeckel2007PRD} as well as by Steigman \cite{Steigman2013PRD} to obtain the constraints on the MeV asymmetric dark matter with the effective number of neutrinos $N_{\rm eff}$ at big bang nucleosynthesis.

If an extra particle is light enough, this light particle has contributed directly to the effective number of neutrinos as the so-called dark radiation (direct contribution case). From the requirement of $\rho_\chi \le \rho_{\rm DR}$, we have obtained the upper limit on the asymmetric dark matter number asymmetry $\epsilon_{\rm max} \sim 0.01$ for $\Delta N_\nu \lesssim 2.67$.

Although the extra particles are not light enough to contribute directly to the effective number of neutrinos, its annihilation yields either increase or decrease the effective number of neutrinos (indirect contribution case). If the MeV asymmetric dark matter couples more strongly to the neutrinos than to the photons and electrons ($\Gamma_{\chi\gamma, \chi e} \gg \Gamma_{\chi\nu}$) or if the MeV asymmetric dark matter couples more strongly to the electrons and photons than to the neutrinos ($\Gamma_{\chi\gamma, \chi e} \ll \Gamma_{\chi\nu}$), the constraint on $m_\chi^{\rm min}$ with $N_{\rm eff}$ is obtained in Eq.(\ref{Eq:constraint_m_chi_min_indirect}). For example,
\begin{eqnarray}
m_\chi^{\rm min} \simeq 18 \ {\rm MeV},
\end{eqnarray}
if the effective number of neutrinos is just the standard value $N_{\rm eff}=3.0$ as shown in Eq.(\ref{Eq:constraint_m_chi_min_indirect_neff3}). The constraint on $\epsilon$ with $N_{\rm eff}$ is also obtained in Eq.(\ref{Eq:constraint_epsilon_max_indirect}). For example $\epsilon_{\rm max} \simeq 10^{-7}$ for $N_{\rm eff}=3.0$ as shown in Eq.(\ref{Eq:constraint_epsilon_max_indirect_neff3}).

If the origin of the observed 511 keV gamma-ray is the annihilating MeV asymmetric dark matter and/or the all of cold dark matter is made of the MeV asymmetric dark matter, $\epsilon \simeq 1.0\times 10^{-8}$ is consistent with data. The constraint on $\epsilon$ is not a strong constraint in both of direct ($\epsilon_{\rm max} \sim 0.01$, for $\Delta N_\nu \lesssim 2.67$) and indirect contribution cases ($\epsilon_{\rm max} \simeq 10^{-7}$ for $N_{\rm eff}=3.0$).

On the other hand, the lower limit of the asymmetric dark matter mass $m_\chi^{\rm min} \simeq 18$ MeV for $N_{\rm eff} \simeq 3.0$ in the indirect contribution case is strict constraint. From the galactic 511 keV gamma-ray line observation, the symmetric dark matter mass may be less than about a few 10 MeV (or a few MeV). We can expect that the range of the asymmetric (as well as symmetric) MeV dark matter mass is so narrow to satisfy $m_\chi \simeq 18$ MeV in the case of $\Gamma_{\chi\gamma, \chi e} \gg \Gamma_{\chi\nu}$ or $\Gamma_{\chi\gamma, \chi e} \ll \Gamma_{\chi\nu}$. The constraint may be useful to check or construct the MeV asymmetric dark matter models.

The role of the number asymmetry of unobserved MeV scale particles is not strong at BBN. The case of ADM is not significantly different from the case of symmetric DM without large chemical potential; however, the explicit considerations of the relation between the chemical potential of ADM and effective number of neutrinos at BBN are shown for the first time. Our actual calculations are important complement to any previously reported results in the literature. The method to discuss the dependence of the chemical potential of ADM on the effective number of neutrinos give a little useful step toward solving puzzles in the dark matter problems. 




\begin{thebibliography}{99}
\bibitem{Zurek2014PREP}
K. M. Zurek, \Journal{\PREP}{537}{91}{2014}.

\bibitem{Gelmini1987NPB}
G. Gelmini, L. I. Hall and M. Lin, \Journal{\NPB}{281}{726}{1987}.

\bibitem{Falkowski2011JHEP}
A. Falkowski, J. T. Ruderman and T. Volansky, \Journal{\JHEP}{05}{106}{2011}.

\bibitem{DAgnolo2015PRD}
R. T. D'Agnolo and A. Hook, \Journal{\PRD}{91}{115020}{2015}.

\bibitem{Izaguirre2015ArXiv}
E. Izaguirre, G. Krnjaic and M. Pospelov, \Journal{\PRD}{92}{095014}{2015}.

\bibitem{Steigman2012AHEP}
G. Steigman, Adv. High Energy Phys. id:268321 (2012).

\bibitem{Sorensen2013PASA}
S. R.-S{\o}rensen, D. Parkinson and T. M. Davis, PASA {\bf 30}, e029 (2013).

\bibitem{Lesgourgues2013}
J. Lesgourgues, G. Mangano, G. Miele and S. Pastor, ``Neutrino Cosmology", Cambridge University Press (2013).

\bibitem{Mangano2005NPB}
G. Mangano, G. Miele, S. Pastor, T. Pinto, O. Pisanti and P. D. Serpico, \Journal{\NPB}{729}{221}{2005}.

\bibitem{Kolb1986PRD}
E. W. Kolb, M. S. Turner and T. P. Walker, \Journal{\PRD}{34}{2197}{1986}.

\bibitem{Serpico2004PRD}
P. D. Serpico and G. G. Raffelt, \Journal{\PRD}{70}{043526}{2004}.

\bibitem{Boehm2012JCAP}
C. B\oe hm, M. J. Dolan and C. McCabe, \Journal{\JCAP}{12}{027}{2012}.

\bibitem{Ho2013PRD1}
C. M. Ho and R. J. Scherrer, \Journal{\PRD}{87}{023505}{2013}.

\bibitem{Ho2013PRD2}
C. M. Ho and R. J. Scherrer, \Journal{\PRD}{87}{065016}{2013}.

\bibitem{Jacques2013PRD}
T. D. Jacques, L. M. Krauss and C. Lunardini, \Journal{\PRD}{87}{083515}{2013}.

\bibitem{Valentino2013JCAP}
E. D. Valentino, A. Melchiorri and O. Mena, \Journal{\JCAP}{11}{018}{2013}.

\bibitem{Franca2013PRD}
U. Fran{\c{c}}a, R. A. Lineros, J. Palacio and S. Pastor, \Journal{\PRD}{87}{123521}{2013}.

\bibitem{Steigman2013PRD}
G. Steigman, \Journal{\PRD}{87}{103517}{2013}.

\bibitem{Nollett2014PRD}
K. M. Nollett and G. Steigman, \Journal{\PRD}{89}{083508}{2014}.

\bibitem{Mirizzi2015PRD}
A. Mirizzi, G. Mangano, O. Pisanti and N. Saviano, \Journal{\PRD}{91}{025019}{2015}.

\bibitem{Nollett2015PRD}
K. M. Nollett and G. Steigman, \Journal{\PRD}{91}{083505}{2015}.

\bibitem{Buen-Abad2015PRD}
M. A. Buen-Abad, G. M.-Tavares and M. Schmaltz, \Journal{\PRD}{92}{023531}{2015}.

\bibitem{Heo2015arXiv}
J. H. Heo and C. S. Kim, arXiv:1504.00773v2 (2015).

\bibitem{Nussinov1985PL}
S. Nussinov, \Journal{\PLBOLD}{165B}{55}{1985}.

\bibitem{Barr1990PLB}
S. M. Barr, R. S. Chivukula and E. Farhi, \Journal{\PLB}{241}{387}{1990}.

\bibitem{Kaplan1992PRL}
D. B. Kaplan, \Journal{\PRL}{68}{741}{1992}.

\bibitem{Kaplan2009PRD}
D. E. Kaplan, M. A. Luty and K. M. Zurek, \Journal{\PRD}{79}{115016}{2009}.

\bibitem{Graesser2011JHEP}
M. L. Graesser, I. M. Shoemaker and L. Vecchi, \Journal{\JHEP}{10}{110}{2011}.

\bibitem{Iminniyaz2011JCAP}
H. Iminniyaz, M. Drees and X. Chen, \Journal{\JCAP}{07}{003}{2011}.

\bibitem{Lin2012PRD}
T. Lin, H.-B. Yu and K. M. Zurek, \Journal{\PRD}{85}{063503}{2012}.

\bibitem{Boeckel2007PRD}
T. Boeckel and J. Schaffner-Bielich, \Journal{\PRD}{76}{103509}{2007}.

\bibitem{Blennow2012JCAP}
M. Blennow, E. F. Martinez, O. Mena, J. Redondo and P. Serra, \Journal{\JCAP}{07}{022}{2012}.

\bibitem{Scherrer1986PRD}
R. J. Scherrer and M. S. Turner, \Journal{\PRD}{33}{1585}{1986}. Erratum.\Journal{\PRD}{34}{3263}{1986}.

\bibitem{Dolgov1993NPB}
A. D. Dolgov and K. Kainulainen, \Journal{\NPB}{402}{349}{1993}.

\bibitem{Ellwanger2012JCAP}
U. Ellwanger and P. Mitropoulos, \Journal{\JCAP}{07}{024}{2012}.

\bibitem{Gelmini2013JCAP}
G. B. Gelmini, J.-H. Huh and T. Rehagen, \Journal{\JCAP}{08}{003}{2013}.

\bibitem{Baldes2014PRL}
I. Baldes, N. F. Bell, K. Petraki and R. R. Volkas, \Journal{\PRL}{113}{181601}{2014}.

\bibitem{Bell2015PRD}
N. F. Bell, S. Horiuchi and I. M. Shoemaker, \Journal{\PRD}{91}{023505}{2015}.

\bibitem{Steigman1977PLBOLD}
G. Steigman, D. N. Schramm and J. E. Gunn, \Journal{\PLBOLD}{66B}{202}{1977}.

\bibitem{Hut1977PLBOLD}
P. Hut, \Journal{\PLBOLD}{69B}{85}{1977}.

\bibitem{Srednicki1988NPB}
M. Srednicki, R. Watkins and K. A. Olive, \Journal{\NPB}{310}{693}{1988}.

\bibitem{Gondolo1991NPB}
P. Gondolo and G. Gelmini, \Journal{\NPB}{360}{145}{1991}.

\bibitem{Kolb1990}
E. W. Kolb and M. S. Turner, ``The Early Universe", Addison-Wesley (1990).

\bibitem{Wantz2010PRD}
O. Wantz and E. P. S. Shellard, \Journal{\PRD}{82}{123508}{2010}.

\bibitem{PDG2014}
K. A. Olive, et.al., (Particle Data Group), \Journal{\CPC}{38}{090001}{2014}.

\bibitem{Planck2015arXiv}
P.  A. R. Ade, et. al. (Planck Collaboration), arXiv:1502.01589v2 (Feb,2015).

\bibitem{Kolb1980NPB}
E. W. Kolb and S. Wolfram, \Journal{\NPB}{172}{224}{1980}.

\bibitem{Press1992}
W. H. Press, S. A. Teukolsky, W. T. Vetterling and B. P. Flannery, ``Numerical recipes in C 2$^{nd}$ edition", Cambridge University Press, (1992).

\bibitem{Schwarz2009JCAP}
D. J. Schwarz and M. Stuke, \Journal{\JCAP}{11}{025}{2009}.

\bibitem{Iizuka2015MPLA}
J. Iizuka and T. Kitabayashi, \Journal{\MPLA}{30}{1550003}{2015}.

\bibitem{Ichimasa2014PRD}
R. Ichimasa, R. Nakamura, M. Hashimoto and K. Arai, \Journal{\PRD}{90}{023527}{2014}.

\bibitem{Ichikawa2003PLB}
K. Ichikawa and M. Kawasaki, \Journal{\PLB}{570}{154}{2003}.

\bibitem{Boehm2013JCAP}
C. B\oe hm, M. J. Dolan and C. McCabe, \Journal{\JCAP}{08}{041}{2013}.

\bibitem{Boehm2004PRL}
C. B\oe hm, D. Hooper, J. Silk, M. Casse and J. Paul, \Journal{\PRL}{92}{101301}{2004}.

\bibitem{Fayet2004PRD}
P. Fayet, \Journal{\PRD}{70}{023514}{2004}.

\bibitem{Beacom2005PRL}
J. F. Beacom, N. F. Bell and G. Bertone, \Journal{\PRL}{94}{171301}{2005}.

\bibitem{Boehm2006PLB}
C. B\oe hm, J. Orloff and P. Salati, \Journal{\PLB}{641}{247}{2006}.

\bibitem{Jacoby2007JHEP}
C. Jacoby and S. Nussinov, \Journal{\JHEP}{05}{017}{2007}.

\bibitem{Pospelov2007PLB}
M. Pospelov and A. Ritz, \Journal{\PLB}{651}{208}{2007}.

\bibitem{Finkbeiner2007PRD}
D. P. Finkbeiner and N. Weiner, \Journal{\PRD}{76}{083519}{2007}.

\bibitem{Huh2008PRD}
J.-H. Huh, J. E. Kim, J.-C. Park and S. C. Park, \Journal{\PRD}{77}{123503}{2008}.

\bibitem{Lingenfelter2009PRL}
R. E. Lingenfelter, J. C. Higdon and R. E. Rothschild, \Journal{\PRL}{103}{031301}{2009}.

\bibitem{Vincednt2012JCAP}
A. C. Vincent, P. Martine and J. M. Cline, \Journal{\JCAP}{04}{022}{2012}.

\bibitem{Huang2013PRD}
J. Huang and A. E. Nelson, \Journal{\PRD}{88}{033016}{2013}.

\bibitem{Graham2015PRD}
P. W. Graham, S. Rajendran, K. V. Tilburg and T. D. Wiser, \Journal{\PRD}{91}{103524}{2015}.

\bibitem{Patra2015PRD}
S. Patra, N. Sahoo and N. Sahu, \Journal{\PRD}{91}{115013}{2015}.

\bibitem{Arcadi2015JCAP}
G. Arcadi, L. Covi and F. Dradi, \Journal{\JCAP}{07}{023}{2015}.

\bibitem{Cline2012PLB}
J. M. Cline and A. R. Frey, \Journal{\PLB}{706}{384}{2012}.

\bibitem{Masina2012JCAP}
I. Masina, P. Panci and F. Sannino, \Journal{\JCAP}{12}{002}{2012}.

\bibitem{Giesen2015ArXiv}
G. Giesen, M. Boudaud, Y. G\'{e}nolini, V. Poulin, M. Cirelli, P. Salati and P. D. Serpico, \Journal{\JCAP}{09}{023}{2015}.

\bibitem{McDermott2015PDU}
S. D. McDermott, \Journal{\PDU}{7-8}{12}{2015}.

\bibitem{Regis2015PRL}
M. Regis, J.-Q. Xia, A. Cuoco, E. Branchini. N. Fornengo and M. Viel, \Journal{\PRL}{114}{241301}{2015}.

\bibitem{Sameth2015PRL}
A. G.-Sameth, M. G. Walker and S. M. Koushiappas, \Journal{\PRL}{115}{081101}{2015}.

\bibitem{McDonald2000PRD}
P. McDonald, R. J. Scherrer and T. P. Walker, \Journal{\PRD}{63}{023001}{2000}.

\bibitem{Ali-Haimound2015PRL}
Y. A.-Ha\"{\i}mound, J. Chluba and M. Kamionkowski, \Journal{\PRL}{115}{071304}{2015}.

\bibitem{DAgnolo2015PRL}
R. T. D'Agnolo and J. T. Ruderman, \Journal{\PRL}{115}{061301}{2015}.

\bibitem{Steigman2015PRD}
G. Steigman, \Journal{\PRD}{91}{083538}{2015}.

\bibitem{Frieman1990PRD}
J. A. Frieman, E. W. Kolb and M. S. Turner, \Journal{\PRD}{41}{3080}{1990}.

\bibitem{BeacomPRL2006}
J. F. Beacom and H. Y\"{u}ksel, \Journal{\PRL}{97}{071102}{2006}.

\bibitem{Sizun2006PRD}
P. Sizun, M Cass\'{e} and S. Schanne, \Journal{\PRD}{74}{063514}{2006}.

\bibitem{Boehm2004NPB}
C. B{\oe}hm and P. Fayet, \Journal{\NPB}{683}{219}{2004}.

\end{thebibliography}
\end{document}